\documentclass[fleqn]{2026SCGE}
\usepackage{bm}
\usepackage{algorithm}
\begin{document}
\ensubject{subject}

\ArticleType{Article}
\SpecialTopic{SPECIAL TOPIC: }
\Year{2026}
\Month{January}
\Vol{69}
\No{1}
\DOI{??}
\ArtNo{000000}
\ReceiveDate{March 11, 2026}
\AcceptDate{June 24, 2023}

\title{First Observational Evidence for Split Infall Flow of Cosmic Filaments into Clusters}

\author[1,2]{Ji Yao}{ji.yao@shao.ac.cn}

\author[1,2]{Huanyuan Shan}{hyshan@shao.ac.cn}

\author[3,4,5]{Pengjie Zhang}{}

\author[3,4,5]{Xiaohu Yang}{}

\author[3]{Jiale Zhou}{}

\author[3,4,5]{Jiaxin Han}{}

\author[1,2]{\newline\\Peng Wang}{}

\author[1]{Haojie Xu}{}

\address[1]{Shanghai Astronomical Observatory (SHAO), Chinese Academy of Sciences, Nandan Road 80, Shanghai, 200030, China}

\address[2]{University of Chinese Academy of Sciences, Beijing, 101408, China}

\address[3]{Department of Astronomy, School of Physics and Astronomy, Shanghai Jiao Tong University, Shanghai, 200240, China}

\address[4]{Key Laboratory for Particle Astrophysics and Cosmology (MOE)/Shanghai Key Laboratory for Particle Physics and Cosmology, Shanghai, 200240, China}

\address[5]{Tsung-Dao Lee Institute, Shanghai Jiao Tong University, Shanghai, 200240, China}


\abstract{Velocity fields in the cosmic web are fundamental to structure formation but remain difficult to observe directly beyond the linear regime. Here we present observational evidence that galaxy filaments connecting pairs of galaxy clusters undergo a split infall, with opposite velocity flows toward the two clusters. Using spectroscopic galaxies from the Sloan Digital Sky Survey, we isolate the internal filament velocity field by subtracting its rigid-body background motion and Hubble flow, and detect this effect at a significance greater than $5\sigma$ across a wide range of cluster and filament selections. The measured velocity profile exhibits a sign reversal near the filament midpoint and a maximum infall amplitude of $\sim30$ km/s ($\sim20$ km/s projected onto the line-of-sight) for clusters of mass $\sim10^{14.3}M_\odot$, substantially lower than expected for infall from an average cosmic environment. Multiple results on density-velocity correlation, mass-dependency, and validation with simulations indicate that filaments dynamically respond to competing gravitational potentials rather than acting as passive mass transport channels. Our results establish a new observational window on quasi-linear velocity fields in the cosmic web and provide a promising probe of mass measurement, testing gravity and velocity reconstruction with upcoming wide-field spectroscopic surveys.}


\maketitle


\begin{multicols}{2}
\section{Introduction}\label{sec1}

The velocity field is a fundamental probe of cosmic structure formation, encoding the growth of density perturbations and the action of gravity across both linear and non-linear regimes \cite{Zhang2007,Hahn2015}. It governs matter transport within the cosmic web \cite{Cautun2014}, the assembly of dark matter halos and galaxy clusters \cite{Fong2021,Zhou2023,Zhou2025,Sheth2001,Jing1998}, and the dynamics of filaments connecting these dense nodes \cite{Libeskind2015,Wang2021}, as well as galaxy and gas motions on smaller scales \cite{Li2021,Faucher-Giguere2023}. Despite its central role, direct observational access to velocity fields beyond the linear regime remains limited. In galaxy surveys, redshifts combine contributions from cosmic expansion and peculiar motions, making it difficult to isolate velocities without additional modeling \cite{cosmo-review2022}. Velocity reconstruction techniques infer velocity fields from density maps under assumptions of gravitational instability \cite{Wang2012,Yu2017,Yu2019,Xiao2025}, while the kinetic Sunyaev-Zel'dovich effect probes electron motions through secondary cosmic microwave background anisotropies \cite{Chen2022,Chen2023,Zheng2013,Li2024}. However, both approaches face practical and conceptual limitations, particularly on quasi-linear and non-linear scales \cite{Stiskalek2025,Aghanim2001}.

\textcolor{black}{Filaments are elongated, anisotropic structures \cite{1996Natur.380..603B, 2018MNRAS.473.1195L} with characteristic widths \cite{2024MNRAS.532.4604W}, density \cite{2020A&A...641A.173G} and kinematic \cite{Wang2021} contrasts that influence the spatial \cite{ 2007MNRAS.375..489H, 2013ApJ...779..160Z}, kinematic \cite{2009ApJ...706..747Z} and evolutionary properties \cite{2018MNRAS.473.1562W} of galaxies and halos. Filaments therefore act as channels of mass and angular momentum transfer into clusters, with their properties tightly linked to their connected clusters \cite{Wang2021,2024MNRAS.532.4604W}.} Regarding structure-specific velocity fields, extensive research has characterized the infall patterns and angular momentum of clusters and filaments \cite{Wang2021,Diaferio1997,Libeskind2015}, as well as their complex correlations with underlying density fields and host halo properties \cite{Zhou2023,Zhou2025,Fong2021,More2015,Gabriel-Silva2025,Hahn2007,Codis2015,Yang2021,Yang2005,Yang2005group}. However, the velocity interaction between clusters and filaments, which is a critical part of cosmic structure formation, remains largely unexplored through direct observation \cite{Tang2025}.

In this work, we investigate the internal velocity field of filaments in cluster-filament-cluster systems identified in the Sloan Digital Sky Survey (SDSS) galaxy group catalogs \cite{Tempel2017,Eisenstein2011,Alam2015}, and test whether filamentary flows exhibit a split infall pattern under the competing gravitational influence of clusters on both sides. By exploiting the geometry of cluster pairs, we can resolve velocity flows along different segments of the filaments and directly measure velocity gradients on quasi-linear scales. This provides complementary information to density-based velocity reconstruction, which primarily probes the linear regime. It also provides direct insight into how matter is supplied to galaxy clusters through the cosmic web \cite{Han2018}.

\section{Methods}\label{Sec method}
\subsection{Observational Data}

Observationally, it is difficult to directly see how a filament falls into galaxy clusters, while this effect is expected in both gravitational theory and cosmological simulations. The key problem is the degeneracy between Hubble flow and peculiar velocity: in galaxy surveys, the observed redshift of a single galaxy contains cosmological redshift due to Hubble flow \footnote{$v=H_0 r$, two objects separated by distance $r$ are effectively moving apart at velocity $v$ controlled by the Hubble constant $H_0$} and Doppler redshift due to peculiar velocity. This problem can be reconsidered when a filament exists between a pair of clusters, as in this work we propose a simple design to find how the filament falls separately into the two clusters, which includes: the speed of the infall flow, its behavior, and its dependence on mass.

The components that build up the redshift distribution of a filament are resolved as: Hubble flow, rigid-body motions including translation and rotation, a dynamical flow motion on top of the rigid-body, and pairwise velocity of the two clusters as a background that linearly stretches or compresses the filament. We find that redshift introduced by each term, except the dynamical flow of the filament itself, scales as $z \propto v \propto r$, where $v$ here denotes different types of velocity, and $r$ denotes the distance to the close cluster projected onto the filament direction. Therefore, one can linearly interpolate a ``rigid-body background'' redshift at each point on the filament based on the two clusters at two ends, and subtract this background redshift, effectively subtracting the Hubble flow while fixing the velocity of the two clusters at zero. Consequently, the excess redshift $\Delta z$ after subtracting the rigid-body background motion will come from the flow motion on the filament. 

In this work, we use SDSS spectroscopic galaxies and galaxy groups \cite{Tempel2017} to select clusters, filaments, and measure the flow motion. The methodology to extract the flow motion is straightforward: \\
\textbf{(1)} In comoving space, we select a cluster pair (cluster A and B) that is sufficiently close ($2<L<6$ Mpc, but not too close so that merging clusters and possible cross-contamination are removed), and search for galaxies residing in a cylindrical space (width $W<1.2\times r_{200c}$ where $r_{200c}$ is the radius of the larger cluster) with the two clusters centered at its ends. These galaxies will represent the connecting filament between the cluster pair.\\
\textbf{(2)} Define the filament axis as the straight line connecting the two clusters, and project each galaxy at location $P$ with observed redshift $z_i$ (for the i-th galaxy) onto the straight line to find its normalized location $x_i=\vec{AP}\cdot\vec{AB}/\vec{AB}^2$ ($0<x_i<1$) on the filament, with the close cluster A at location $x_A=0$ and the distant cluster B at location $x_B=1$.\\
\textbf{(3)} Linearly interpolate the redshift at each point on the line ($z(x)=kx+b$) based on the two clusters (cluster A with $x_A$, $z_A$ and cluster B with $x_B$, $z_B$) at two ends. This will represent the redshift introduced by position (i.e. Hubble flow) and velocity along the line-of-sight direction, including translation and rotation as a cluster-cluster pair.\\
\textbf{(4)} By subtracting the interpolated redshift in step (3) from the redshift of galaxies in step (2), one obtains the excess redshift $\Delta z_i=z_i-z(x_i)$, representing the $flow$ motion beyond the rigid-body background motion in the filament frame. We note in this way the two clusters will have $\Delta z=0$, so that their pairwise velocity is also removed.\\
\textbf{(5)} Finally, considering a single filament does not contain enough galaxies to clearly describe the velocity flow, we choose to stack multiple filaments that satisfy the following criteria: cluster richness $\lambda_c>5$, mass $M_{200c}>10^{13.5}M_\odot$, filament richness $1<\lambda_f<5$, length $2<L<6$ Mpc, width $W<1.2\times r_{200c}$.

\begin{figure*}[t!]
	\centering
	\includegraphics[width=0.8\textwidth]{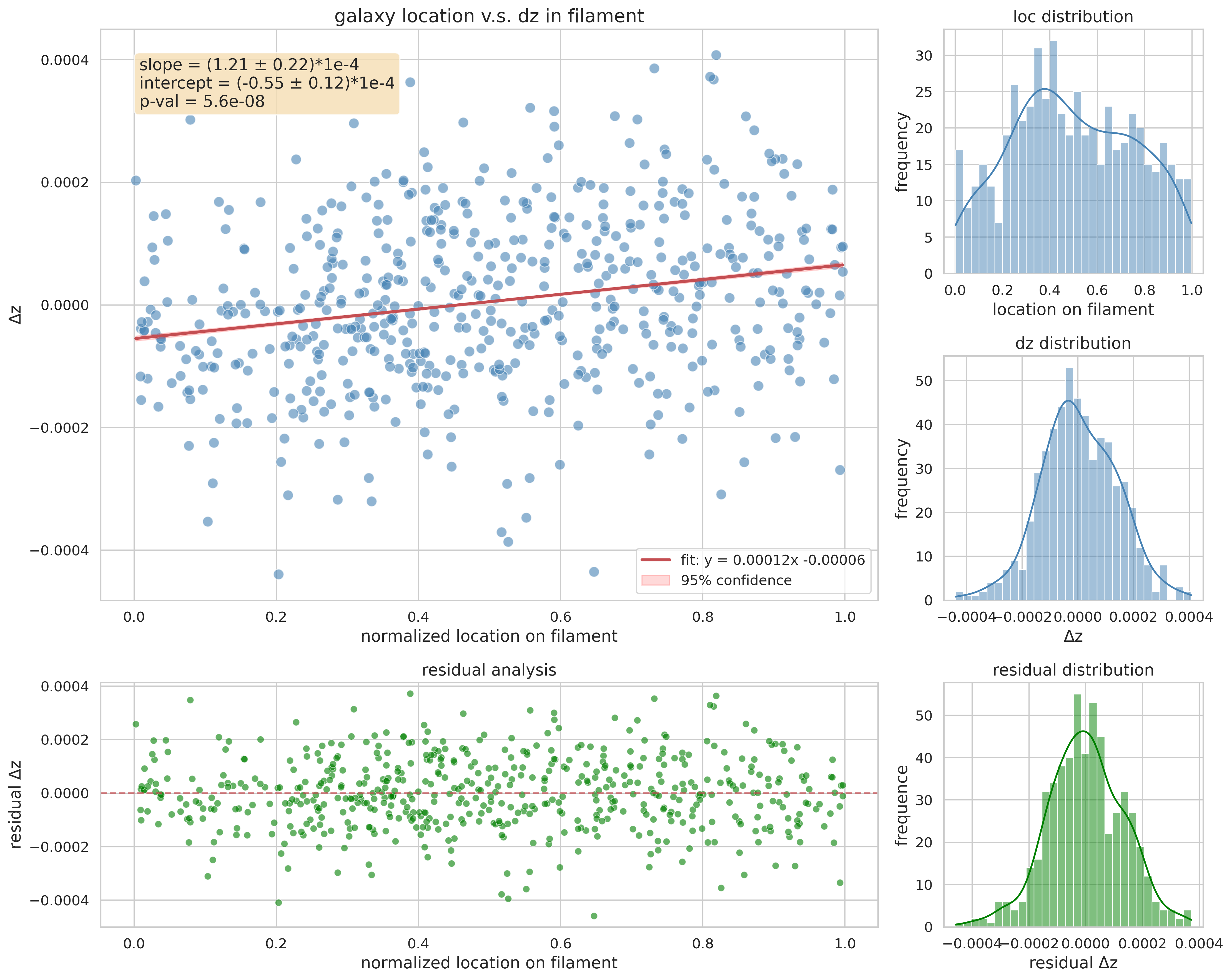}
	\caption{Distribution of the excess redshift of galaxies $\Delta z$ after linearly subtracting rigid-body background redshift of the filament, representing the flow motion on the filament. The top-left panel shows the excess redshift $\Delta z$ v.s. location on the filament, with each (blue) point representing a galaxy. We fit a line (red) to the points and get a slope that deviates from 0 at $5.5\sigma$ significance, providing strong evidence that the filament is flowing from the center to the two sides. The top-right two panels show how the galaxies are distributed on different locations and $\Delta z$. The non-uniform galaxy line-density is not only due to selection effects, but also due to filament dynamics, with further proof presented in Appendix \ref{Appendix RSD Cylinder}. The two bottom panels in green show the residual distribution after subtracting the linear fit.}\label{Fig linear_fit}
\end{figure*}

After the above selections, we obtain 4728 galaxy groups as clusters, and only 360 cluster-filament-cluster systems after the selection of filament length, width, and richness. We stack the filaments by putting the close cluster at $x=0$ and the distant cluster at $x=1$. After linearly removing the rigid-body background motion and cluster pairwise velocity, we successfully obtain the excess redshift $\Delta z_i$ data v.s. normalized location $x_i$ in Fig.\,\ref{Fig linear_fit}. Under the fiducial selection, the stacked filaments have the following average properties: length $L=4.7$ Mpc, width $W=1.6$ Mpc, cluster mass $M_{200c}\sim10^{14.3}M_\odot$, filament mass $M_f\sim10^{13.32}M_\odot$, and filament-redshift ($\hat{f},\hat{z}$) alignment angle $\cos\langle\hat{f},\hat{z}\rangle\sim0.52$, with details provided in Appendix \ref{Appendix Filament Properties}. We note that the radius definition $r_{200c}$ and mass definition $M_{200c}$ we use come from the group catalog \cite{Tempel2017} and are based on the halo profile density at 200 times the critical density of the Universe.

\subsection{Model Validation} \label{Sec model validation}

We construct a velocity emulator from an N-body simulation \cite{Fong2021,Zhou2023,Zhou2025} to describe the general infall flow around clusters/halos, and use the emulator to construct a halo-halo-halo model to mimic the cluster-filament-cluster dynamics and compare it with the filament flow in this work. 

The simulation we use is one set from the CosmicGrowth Simulations \cite{Jing2019}, with a P$^3$M code \cite{Jing2002} run under WMAP $\Lambda$CDM cosmology ($\Omega_b=0.0445$, $\Omega_c=0.2235$, $\Omega_\Lambda=0.732$, $h=0.71$, $n_s=0.968$, and $\sigma_8=0.83$). The simulation box size is 600 Mpc/h with $3072^3$ dark matter particles, and the softening length is 0.01 Mpc/h. The halos are processed with Friends-of-Friends (FoF) and then with HBT+ \cite{Han2012,Han2018} to get subhalos and their evolutionary histories. The halo catalog covers a virial mass range $10^{11.5}<M_{\rm vir}[M_\odot/h]<3\times10^{15}$, with the minimum mass corresponding to $\sim500$ dark matter particles.

To construct the emulator, we choose halos from 9 snapshots covering a redshift range $0<z<3$, and 9 mass bins covering $10^{12}<M_{\rm vir}[M_\odot/h]<10^{15}$. We linearly interpolate the velocity profile as a function of $z$, log$_{10}(M_{\rm vir}/M_\odot)$, and comoving distance to the halo center $r$ [Mpc]. The resulting velocity emulator $v(z,M,r)$ and the density emulator $\rho(z,M,r)$ are constructed and shown in Appendix \ref{Appendix Emulator}. Different profiles generated from these emulators are then stacked following the cluster-filament-cluster spatial distribution as our model for comparison with the observation, which we will show later in Sec. \ref{Sec results}.

\textcolor{black}{In addition, since the simulation uses $M_{\rm vir}$ while the observation uses $M_{\rm 200c}$, we use the Core Cosmology Library (CCL) \cite{Chisari2019} to convert different mass definitions. During this process, a concentration-mass (c-M) relation needs to be assumed, and we compare two widely used models: Duffy08 \cite{Duffy2008} and Diemer15 \cite{Diemer2015}. The different choices can lead to a $\sim3\%$ difference in mass conversion, which is negligible. We adopt the Diemer15 c-M relation to convert the observational $M_{\rm 200c}$ to $M_{\rm vir}$ for use in the emulators.}

\section{Results}\label{Sec results}

\begin{figure}[H]
	\centering
	\includegraphics[width=0.5\textwidth]{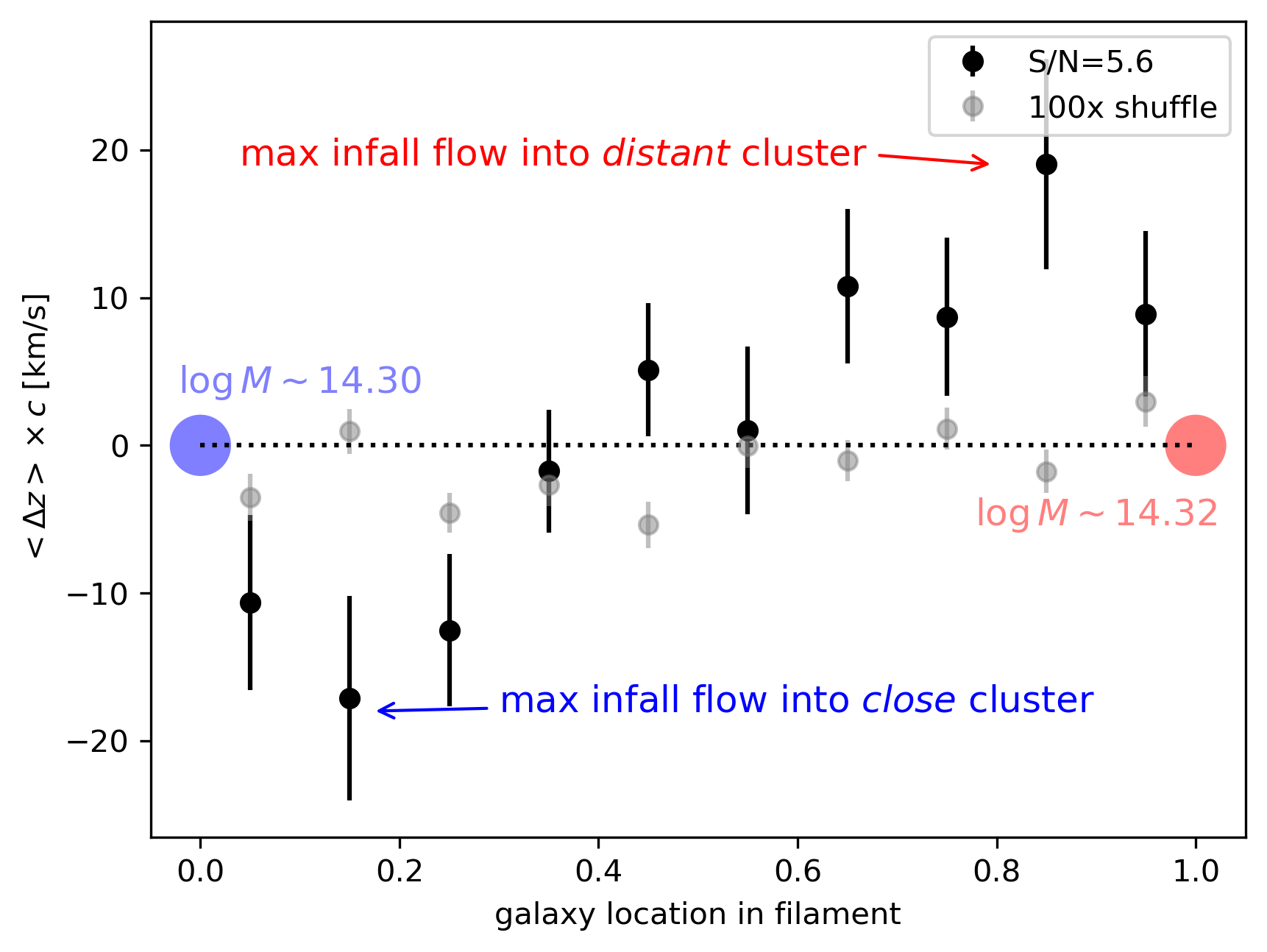}
	\caption{Mean velocity flow measurement along the filament. This figure uses the same data as Fig.\,\ref{Fig linear_fit}, but the data are divides into 10 bins to illustrate the spatial variation of the flow. The blue circle corresponds to the closer cluster (with mass $\sim10^{14.3}M_\odot$) and the red circle represents the more distant cluster (with mass $\sim10^{14.32}M_\odot$). The dotted line represents no velocity flow with respect to the ``rigid-body'' frame of the filament. The black data reject the null hypothesis with a significance of $5.6\sigma$, suggesting the filament is flowing from the center toward the two ends, reaching a maximum velocity at $\sim20$ km/s. \textcolor{black}{The grey data correspond to a null test that keeps the cluster pairs while randomly shuffling the redshifts of all the galaxies and rerunning the whole pipeline.} The deprojected velocity profile along the filament direction and its comparisons with N-body simulation are shown later.}\label{Fig flow}
\end{figure}

We find a non-vanishing flow along the filament direction by stacking 360 filaments defined from the SDSS group catalog (see Sec.\,\ref{Sec method} for details), with the flow-induced excess redshift of each galaxy visualized in Fig.\,\ref{Fig linear_fit}. Relative to the rigid-body background shown as the x-axis, we observe a flow motion with a slope $(1.21\pm0.22)\times10^{-4}$, i.e., a $5.5\sigma$ significance. To better understand how this velocity flow falls into the clusters, we divide the normalized filament length into 10 bins and calculate the mean velocity in each bin. The resulting mean velocity profile is shown in Fig.\,\ref{Fig flow}. This signal rejects the null hypothesis at $5.6\sigma$, which is consistent with the significance of the slope fitting in Fig.\,\ref{Fig linear_fit}. At the close end, the negative excess redshift ($\Delta z<0$) indicates that material within the filament is moving towards us (the observer), while at the far end, it is receding from us ($\Delta z>0$).

Fig.\,\ref{Fig flow} is the binned average data of Fig.\,\ref{Fig linear_fit}, where we divide the filament linearly into 10 bins. The statistical error and covariance in Fig.\,\ref{Fig flow} are calculated with jackknife resampling, treating each cluster-filament-cluster system as an independent subsample. We find that the resulting covariance matrix is strongly diagonal, suggesting the leakage of galaxies from one bin to another is small, which at some level implies the velocity-location contamination from a single redshift (similar to redshift-space distortion, RSD) is not very significant. \textcolor{black}{The S/N (signal-to-noise ratio) is calculated by the deviation of the data vector from zero, i.e., $\sqrt{\chi^2}=\sqrt{\vec{\Delta z}^\top\cdot{\rm Cov_{\Delta z}^{-1}\cdot\vec{\Delta z}}}$. We tested the S/N calculation with the covariance ($5.6\sigma$) or only the errorbar ($5.8\sigma$) as the covariance is generally diagonal.} Since the number of subsamples (360) is significantly larger than the size of the data vector (10), the calibration of the covariance is $\sim3\%$ and is negligible \cite{Wang2020,Hartlap2007}.

\textcolor{black}{As a further validation of the physics discovered in this work, as well as for identifying where the flow signal resides, we keep the cluster pairs unchanged while randomly shuffling the redshifts of all the galaxies. The shuffled galaxy catalog has a size $100\times$ the SDSS galaxy catalog. We present the corresponding $\Delta z$ measurements in Fig.\,\ref{Fig flow} in grey. This null test is generally consistent with zero, suggesting the observed signal does not come from the selection effecst from the selected cluster pairs, boundary effects of the survey, or density-evolution introduced by flux-limited samples, etc.} We additionally tested how the selection functions above (cluster richness $\lambda_c$, cluster mass $M_c$, filament richness $\lambda_f$, filament length $L$ and filament width $W$) affect the flow. The corresponding results are presented in Appendix \ref{Appendix Functions}. We note the result in Fig.\,\ref{Fig flow} is not the optimal choice but rather a conservative selection. By aggressively expanding the filament richness range, one can achieve an S/N$>7\sigma$.

Symmetry considerations allow for a deeper understanding of this flow motion. The motion beyond rigid-body background could arise from three effects:\\
(1) inflow toward the two clusters;\\
(2) rotational flow around an axis offset from the filament;\\
(3) spinning flow around the long axis of the filament.\\
Due to the isotropy required by the cosmological principle (there is no preferred direction in the Universe), when stacking many filament systems, the contributions from filaments with rotation-in and rotation-out flows should cancel out. Therefore, the filament rotational flow described in (2) should not contribute to the signal in Fig.\,\ref{Fig linear_fit} and \ref{Fig flow}. Similarly, the contributions from spin-in and spin-out flows should cancel, so the filament spinning flow described in (3) should also not contribute. Consequently, the observed flow represents the inflow toward the two clusters. By projecting this flow velocity onto the line-of-sight direction (i.e., the z-direction), we obtain the flow-induced excess redshift $\Delta z$ according to: $v_z = v\times\cos\langle\hat{f},\hat{z}\rangle = c\Delta z$, where $v$ is the flow velocity along the filament, $\hat{f}$ is the filament direction pointing from the close cluster to the distant cluster, $\hat{z}$ is the line-of-sight direction, $c$ is the speed of light. \textcolor{black}{An illustration figure of different modes can be found in Appendix \ref{Appendix illustration}.}

We find several physical features of this filament flow:\\
(a) The amplitude of the infall velocity has a trend of increase to a maximum then decrease outward from the cluster center along the filament, which agrees with the depletion radius feature of general dark matter halos in N-body simulations \cite{Fong2021}. The maximum infall velocity toward each cluster with mass $m\sim10^{14.3}M_\odot$ at redshift $z\sim0.059$ reaches $v_{\rm max}\sim30$ km/s, whose projection on the line-of-sight is $v_{\rm z,max}\sim20$ km/s \footnote{Assuming the line connecting cluster A and cluster B is isotropically oriented, a shell integral gives $\bar{v}_{\rm z}=\langle v\cos\theta\rangle=\frac{\int_0^{\pi/2}\int_0^{\pi/2}v \cos\theta \sin\theta d\theta d\phi}{\int_0^{\pi/2}\int_0^{\pi/2}\sin\theta d\theta d\phi}$=$\frac{1}{2}\bar{v}$. We confirm the mean projection effect of $\langle \cos\theta\rangle\sim0.5$ in Appendix \ref{Appendix Filament Properties}. But we note this is the mean value, while the actual value can vary at different locations. A deprojected velocity profile can be found later in Sec \ref{Sec results}.}.\\
(b) The depletion radius, where the infall velocity into the cluster is maximized, is sensitive to the mass distribution in the system. In the observation (Fig.\,\ref{Fig flow}), this position is at $\sim0.15$ of the filament length (2-6 Mpc, 4.7 Mpc on average), so overall $<1$ Mpc from the cluster center. For comparison, in N-body simulation, for a single halo of similar mass in an average-density environment, this position is at $\sim3$ Mpc from the halo center with maximum infall velocity $>400$ km/s after removing Hubble flow (see the velocity emulator in Appendix \ref{Appendix Emulator}). For a simulated halo-halo system in an average cosmological environment (without an intervening filament), this position is at $\sim1.4$ Mpc from the halo center with maximum infall velocity $\sim60$ km/s (later in Sec \ref{Sec results}). This is due to the presence of the filament and the other cluster as a special environment. \\
\begin{figure}[H]
	\centering
	\includegraphics[width=0.45\textwidth]{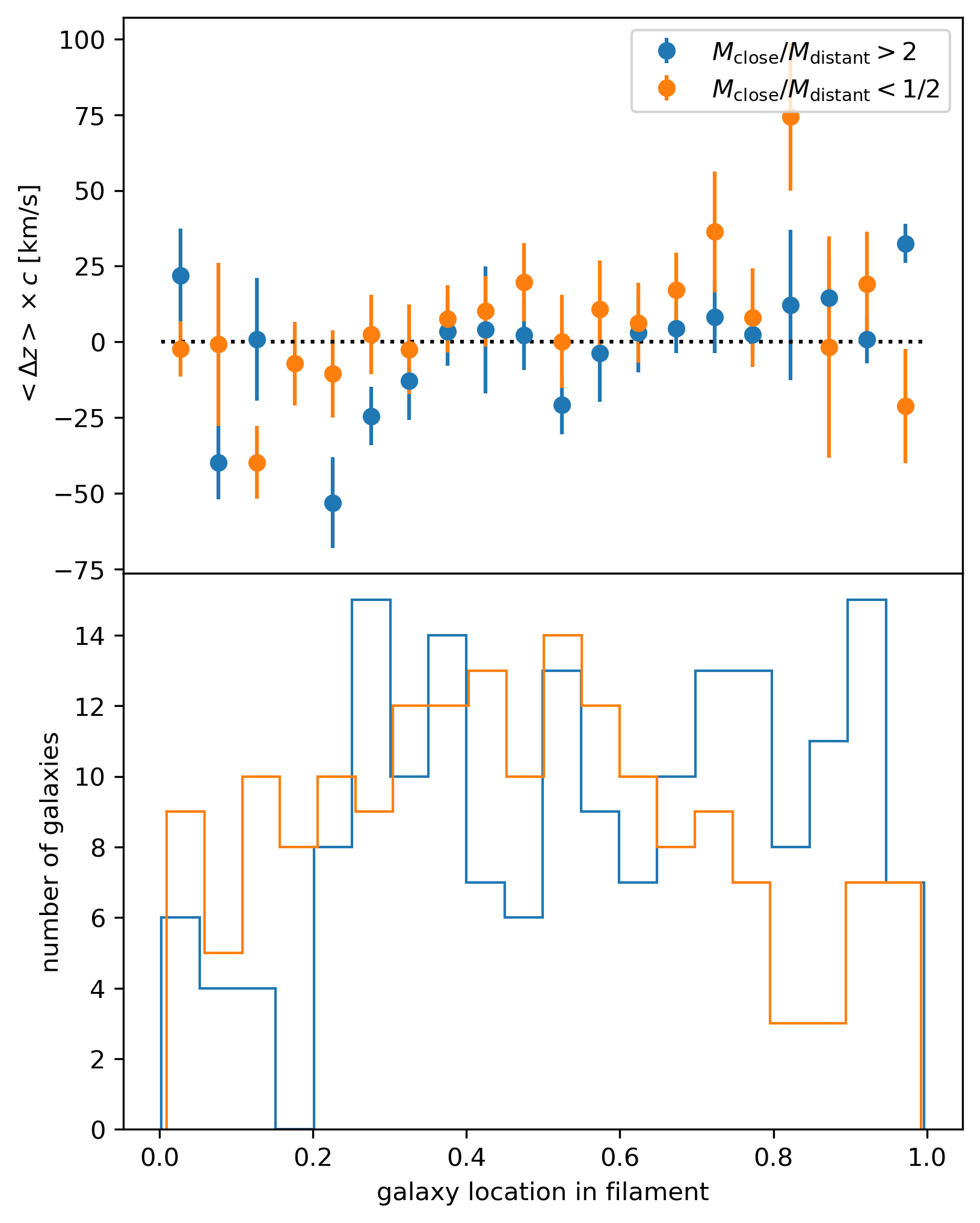}
	\caption{How unbalanced mass affects velocity flow (top) and galaxy distribution (bottom) on the filament. When we further divide the sample according to whether the more massive cluster is at the near end (blue) or at the far end (orange), the filament is generally flowing towards the more massive cluster. Meanwhile, a larger peak velocity ($\sim50$ to $75$ km/s) appears at a larger distance ($\sim0.2$) to the cluster center compared to Fig.\ref{Fig flow}. Significant dips can be found at the location of maximum infall in the galaxy distribution.}\label{Fig unbalanced mass}
\end{figure}
(c) The galaxy line-density is anti-correlated with the amplitude of this flow in the filament. In Fig.\,\ref{Fig linear_fit}, the velocity flow is zero at the filament's midpoint and peaks near the two ends. Meanwhile, the galaxy number count distribution suggests that the density of the filament peaks at the center and decreases at two ends. We demonstrate this density is not due to the redshift-space distortion (RSD) and the cylindrical selection function in Appendix \ref{Appendix RSD Cylinder}, but rather to filament dynamics. This behavior is consistent with the continuity equation, which, in a cylindrical filament approximation, can be expressed as: $d(A\times v\times \rho)/dt=0$, where $A$ denotes the area, $v$ denotes the velocity of the flow, and $\rho$ denotes the density. The density-velocity co-evolution additionally supports the depletion radius as a clear boundary between the cluster and the filament. This point will be further confirmed in the cluster-mass-dependency result that follow.

Besides the symmetry argument discussed previously, further evidence that this flow is driven by the gravity of the cluster pair comes from its mass dependency. In Fig.\,\ref{Fig unbalanced mass}, we see that when the close cluster is at least twice as massive, the excess redshift $\Delta z$ on the filament becomes more negative, indicating that a larger portion of the filament is flowing toward the more massive cluster, and vice versa when the distant cluster is more massive. Under such mass imbalance, both the projected peak infall velocity ($\sim60$ km/s) and the location of depletion radius ($\sim0.2$ of the filament length) increase compared to those in Fig.\,\ref{Fig flow} ($\sim20$ km/s at $\sim0.15$), demonstrating the growing gravitational dominance of the more massive cluster in the system.

\begin{figure*}[t!]
	\centering
	\includegraphics[width=0.8\textwidth]{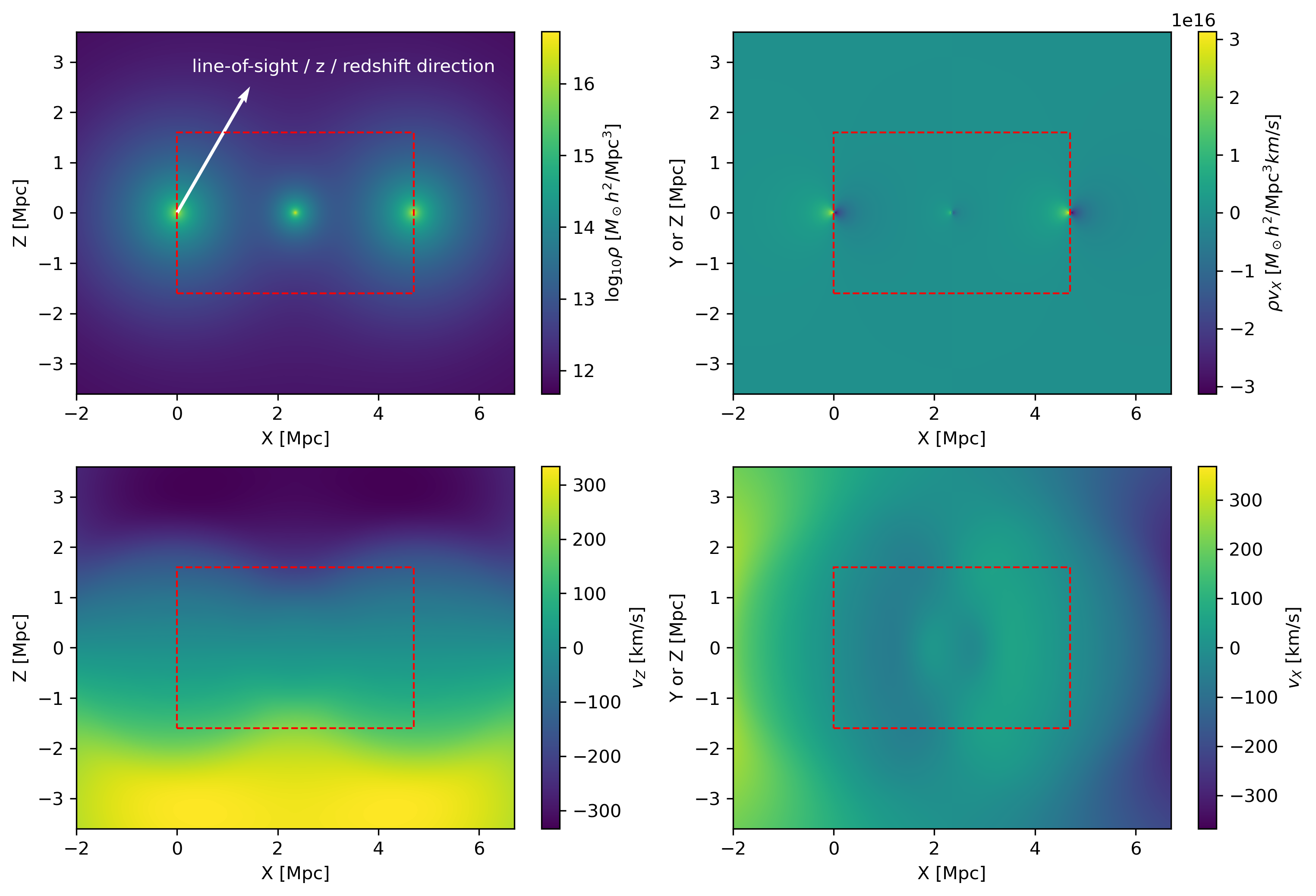}
	\caption{A halo-halo-halo model to mimic the desired cluster-filament-cluster dynamics. The top-left panel shows the spatial configuration of the three halos and their density $\rho$ distribution, with two large halos ($M_{\rm 200c}\sim10^{14.3}M_\odot$) at two ends and a small halo ($M_{\rm 200c}\sim10^{13.32}M_\odot$) in the center. The z-direction (line-of-sight) is $60^\circ$ to the X-axis (filament direction), while Y-axis and Z-axis are perpendicular to X-axis, and z-direction is on the X-Z plane. The top-right panel shows the $\rho v_X$ distribution to describe the mass flow rate. The bottom-right panel shows the distribution of $v_X$, i.e., the velocity along the X-axis (along the filament); its integration over the Y-Z plane within the filament yields the velocity line-profile that we will show later for comparison with the deprojected velocity. The bottom-left panel shows the velocity along the Z-axis $v_Z$. All the panels show the 3D distribution in the central plane ($Y=0$ or $Z=0$) of the filament, while the red dashed box denotes the cylindrical selection function of the filament.}\label{Fig filament model}
\end{figure*}

Then we can build a halo-halo-halo model to mimic the cluster-filament-cluster system we are interested in, using the velocity emulator $v(z,M,r)$ and density emulator $\rho(z,M,r)$ described in Sec. \ref{Sec model validation}. In Fig.\,\ref{Fig filament model} we stack three halos in a straight line, with two large halos of $M_{\rm 200c}\sim10^{14.3}M_\odot$ separated by 4.7 Mpc, and a small halo of $M_{\rm 200c}\sim10^{13.32}M_\odot$ at the center. We define the X-axis as the filament direction, with the Y-axis and the Z-axis perpendicular to the X-axis. The line-of-sight direction, i.e., the z-direction, lies in the X-Z plane and is at an angle of $60^\circ$ to the positive X-axis. We estimate the overall 3D density distribution $\rho_{\rm all}=\sum_{i=1}^{3}\rho_i(X,Y,Z)$ of the system, where $i$ reprecents the i-th halo, and calculate the mass flow rate (MFR) in the X-direction, given by MFR$_X=\sum_{i=1}^{3}\rho_i v_{i,X}$. Then a density-averaged velocity distribution is achieved by $v_X={\rm MFR}_X/\rho_{\rm all}$. We show 3D distributions (on the 2D planes $Y=0$ or $Z=0$, i.e., planes containing the central axis of the filament) of $\rho_{\rm all}$, MFR$_X$, and $v_X$ in Fig.\,\ref{Fig filament model}, where we also present $v_Z$ calculated in a similar way.

\begin{figure}[H]
	\centering
	\includegraphics[width=0.5\textwidth]{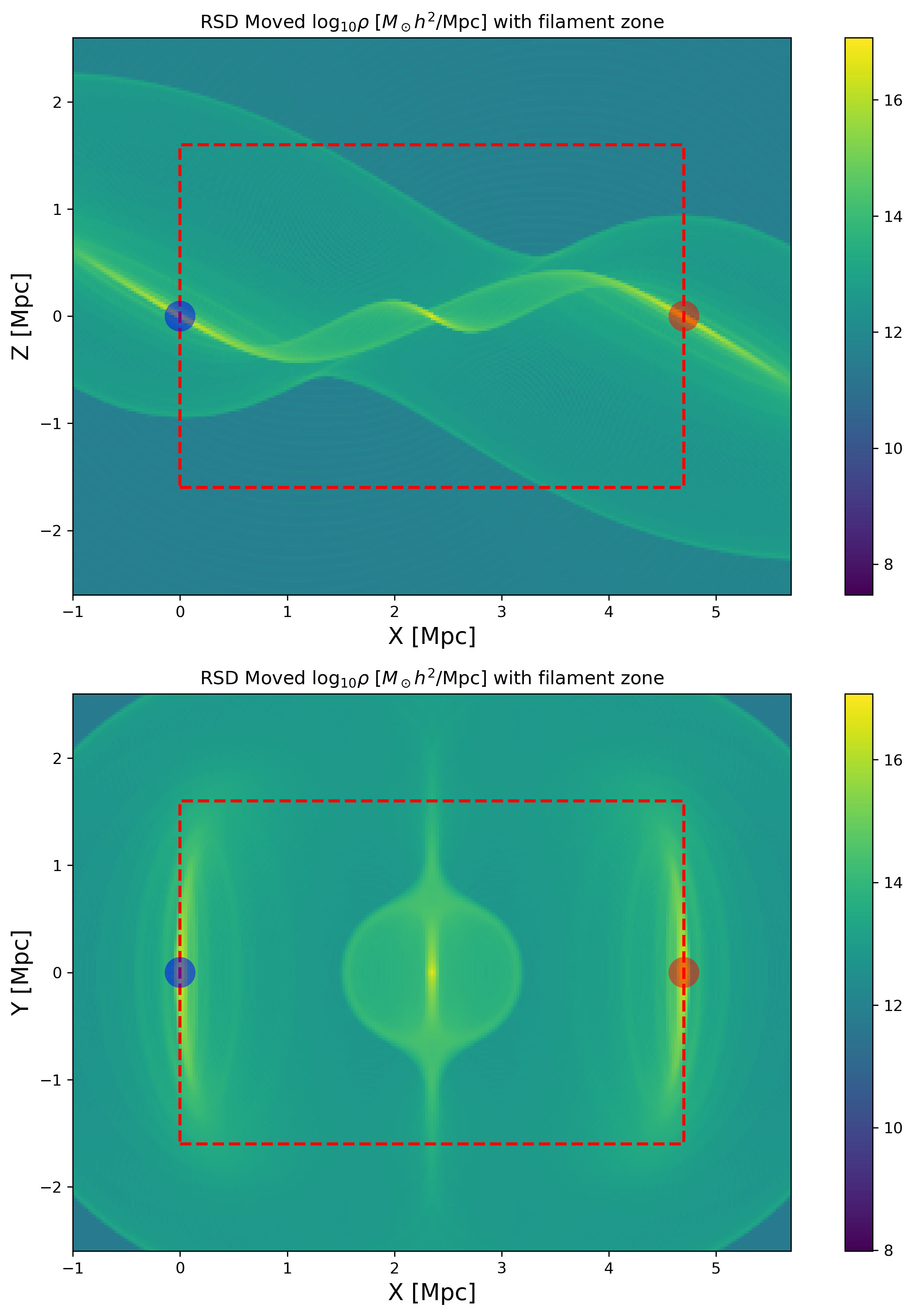}
	\caption{\textcolor{black}{Adding the RSD effect in the model. Based on the density distribution in Fig.\,\ref{Fig filament model} and the associated velocity distribution, we derive the RSD displacement field (see Appendix \ref{Appendix RSD Cylinder}), and the corresponding shifted density. The upper panel shows density in the redshift space in the X-Z plane, and the lower panel shows the density in redshift space in the X-Y plane. The net velocity distribution is also shifted accordingly.}}\label{Fig model add RSD}
\end{figure}

\textcolor{black}{More importantly, we include the redshift-space distortion (RSD) effect \cite{Hamilton1998,Percival2009}, which comes from the degeneracy between cosmological redshift and Doppler redshift $z_{\rm obs}=z_{\rm cosmo}+z_{v}$. Our algorithm in Section \ref{Sec method} directly converts total redshift into cosmological comoving distance, without applying any reconstruction methods. It means the filaments are stacked in redshift-space rather than real-space. Modeling this effect means the 3D density field and velocity field need to be shifted by an additional displacement field; see Appendix \ref{Appendix RSD Cylinder}. We visualize the corresponding density field in Fig.\,\ref{Fig model add RSD} as a comparison to Fig.\,\ref{Fig filament model}. As our model emulates the mean velocity field $v(z,M,r)$, it only contains the Kaiser effect but not the Finger-of-God (FoG) effect. In terms of the measurement, the FoG effect will appear as sample variance and contribute to the statistical errors. In Fig.\,\ref{Fig model add RSD}, the upper panel gives the shape of three stacked Kaiser pancakes, with the line-of-sight direction at an angle of $60^\circ$ to the X-axis, while the lower panel is in the X-Y plane, and thus the peculiar velocity only has an X-component.}

\begin{figure}[H]
	\centering
	\includegraphics[width=0.5\textwidth]{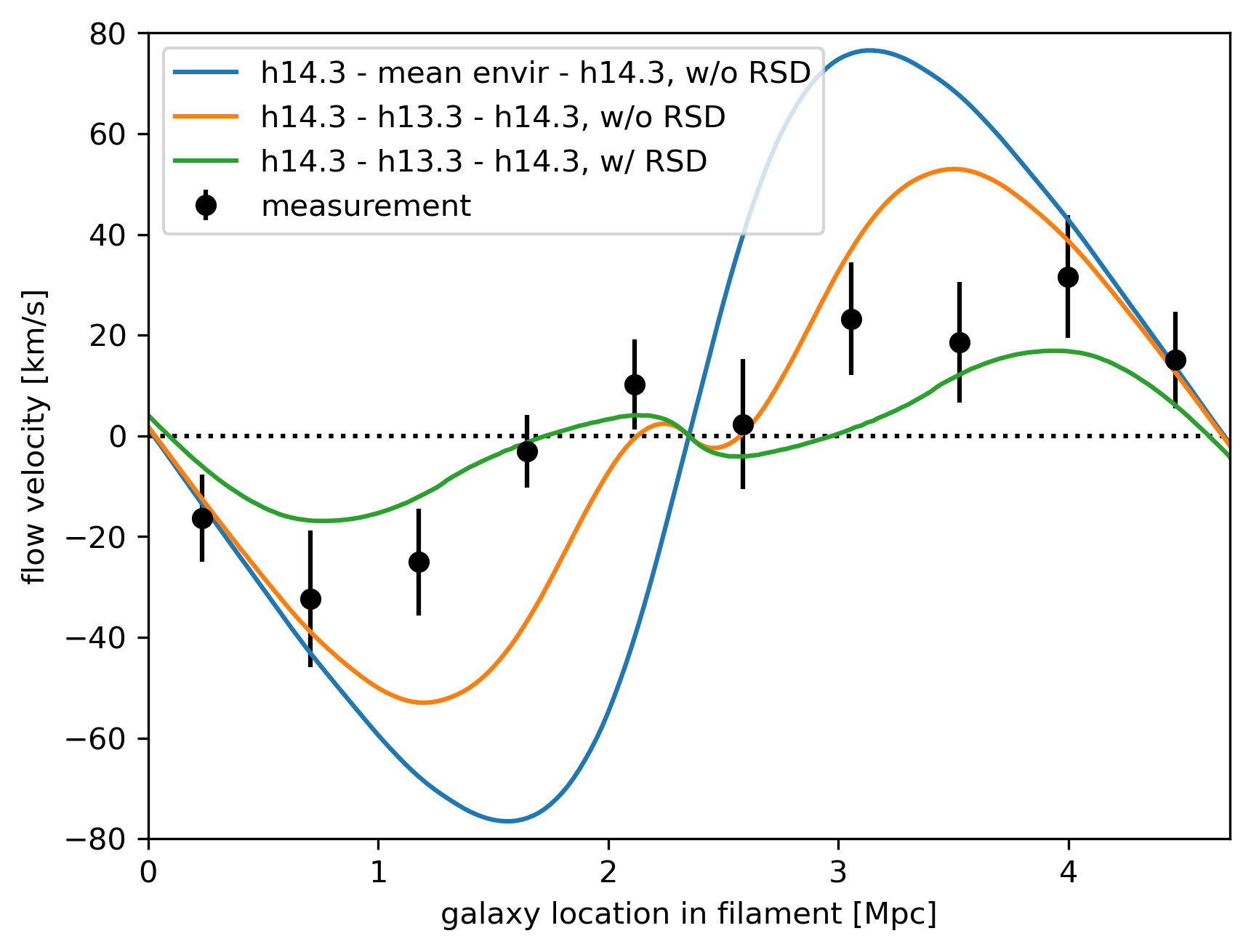}
	\caption{Deprojected velocity flow along the filament direction and comparisons with simulated data using the model from Fig.\,\ref{Fig filament model} and \ref{Fig model add RSD}. The black \textcolor{black}{observational} data show the deprojected velocity flow $v=v_z/\cos\langle\hat{f},\hat{z}\rangle$, using the measurements in Fig.\,\ref{Fig flow}. We see the peak infall velocity is $\sim30$ km/s. The three solid lines correspond to different filament setups, with two $M_{\rm 200c}=10^{14.3}M_\odot$ halos put at the two ends. The blue curve corresponds to the average cosmological environment, with no extra halo to represent the filament at the center. The orange curve includes an extra halo with $M_{\rm 200c}=10^{13.32}M_\odot$ (whose mass corresponds to the optically derived mass of the filament, see Appendix \ref{Appendix Filament Properties}) at the center, similar to Fig.\,\ref{Fig filament model}. \textcolor{black}{The green curve additionally applies the RSD effect in the 3D density field and the velocity field, leading to a suppression in the signal. The green curve agrees with the data in terms of general shape, while the amplitude difference could come from different cosmology and halo definition.}}\label{Fig flow data sim}
\end{figure}

To obtain the 1D flow projected onto the X-axis, similarly to the 3D distributions above, we calculate it as $v_X^{\rm 1D}=\int_Z\int_Y{\rm MFR}_XdYdZ/\int_Z\int_Y\rho_{\rm all}dYdZ$. Meanwhile, the observed flow $v_z=c\Delta z$ in Fig.\,\ref{Fig flow} is the flow projected onto the z-direction, so the flow deprojected onto the filament is $v=v_z/\cos\langle\hat{f},\hat{z}\rangle$. We note this calculation is also performed in each location bin, with jackknife resampling accounting for the variations in the cosine value, similar to the descriptions in Sec.\,\ref{Sec method}. We compare the deprojected observational flow $v$ and the simulated flow $v_X^{\rm 1D}$ in Fig.\,\ref{Fig flow data sim}. \textcolor{black}{In our model, the net flow (blue) peaks at $\sim80$ km/s, which is an order of magnitude smaller than the single-halo situation ($>400$ km/s, see Appendix \ref{Appendix Emulator}) due to the cancellation of the tidal gravitational field. By adding a third halo to represent the filament, the net velocity further reduces (orange) due to the tidal gravitational field of the filament. The RSD effect will further suppress the velocity (green), because our algorithm actually searches a cylindrical shape in redshift-space (Fig.\,\ref{Fig model add RSD}), and galaxies infalling towards the clusters with higher velocity are more likely to be shifted outside the cylinder. Meanwhile, galaxies outside the cylinder in real-space could be shifted into the region in redshift-space, contributing as a negative signal comparing to the desired flow (see the bottom-right panel in Fig.\,\ref{Fig filament model}, where $v_X$ inside and outside the cylinder have opposite signs).}

\textcolor{black}{We see the observation and our model agree well in terms of general shape; however, their overall amplitude could differ at a $\sim50\%$ level. This could be due to four different reasons:\\
	(1) The assumed WMAP cosmology in the simulation is different from the real world. \\
	(2) The halo-halo-halo model we constructed is not a perfect reflection of the cluster-filament-cluster system, as halos are concentrated spherical objects, while the filament should be flatter and elongated along the X-axis. In Appendix \ref{Appendix Filament Model} we present comparisons with more complicated filament mass distributions. \\
	(3) The halo definition could be different, as the observational data \cite{Tempel2017} and simulation data \cite{Zhou2023} we use have different processes to deal with merging halos. This will lead to different density profile $\rho(z,M,r)$ and velocity profile $v(z,M,r)$, and affect whether a galaxy/subhalo is assigned to the cluster or the filament. \\
	(4) The group finder algorithm in observation could introduce biased mass estimation, which could vary from 0.2 dex to 0.45 dex \cite{Sun2022}. A related point is that the group finder we use \cite{Tempel2017} assumes Planck2015 cosmology, which could be biased.}

\section{Summary and Discussion} \label{Sec summary}

Physically, both theory and simulations predict mass flow from less massive regions into more massive ones. In this work, we present a straightforward design to remove Hubble flow with the first measurement of how a filament gets split and falls separately into the clusters at two ends (Fig.\,\ref{Fig flow}). We validate that this flow is gravity-driven via three independent perspectives: 
\\
(1) Observationally, we show the amplitude of the flow is anti-correlated with the galaxy-line-density (Fig.\,\ref{Fig linear_fit}), which is expected from the continuity equation.\\
(2) Also in observation, we show the velocity profile of the flow is sensitive to the mass of the clusters at the two sides (Fig.\,\ref{Fig unbalanced mass}).\\
(3) Theoretically, the measured velocity profile agrees with our model constructed from N-body simulations (Fig.\,\ref{Fig flow data sim}).\\
\textcolor{black}{Additionally, we present three null tests to further validate the observed signal is not due to other effects:\\
(1) We randomly shuffle the redshifts of the galaxies in SDSS and re-run the whole process, and the shuffled result is consistent with zero (Fig.\,\ref{Fig flow}), demonstrating this signal is not due to selection effects but due to large-scale structure, i.e., either clustering or peculiar velocity.\\
(2) In a mock catalog validation, when redshift is only from cosmological redshift, the result is consistent with zero, while after adding peculiar velocity in the mock redshift, a similar signal appears, see Appendix \ref{Appendix mock validation}.\\
(3) We generate 1M random points in a cylinder mimicking our filament properties (z=0.06, L=4.7 Mpc, W=1.6 Mpc, $\cos\langle\hat{f},\hat{z}\rangle=0.5$) and the resulting excess redshift $c\Delta z<0.6$ km/s, see Appendix \ref{Appendix nonlinear d(z)}. This demonstrates the nonlinearity in redshift-distance relation on the filament scale is negligible.\\}
We therefore establish that this flow is gravity-induced, with a significance exceeding $5\sigma$. \textcolor{black}{The most exciting meaning of this flow is not its observation, but a method to break the cosmological-Doppler degeneracy in the redshift of a single galaxy in such a filamentary environment. If the S/N (of the stacked objects) permits, this approach is applicable to get the bulk velocity of the selected structure, and opens a gate for physical discoveries associated with velocity. In terms of the measured flow, some details and their interpretations are discussed as follows.}

This flow delineates a clear boundary between the filament and the cluster, where the infall velocity is maximized (Fig.\,\ref{Fig flow}) and galaxy line-number-density is minimized (Fig.\,\ref{Fig linear_fit}). This feature is similar to the depletion radius of dark matter halos \cite{Fong2021} governed by the continuity equation, while the difference is that the filament here plays the role of a special environment, and the clusters at the two ends slowly ($<30$ km/s) ``steal'' matter from it. 
In addition, we demonstrate that the change in galaxy line-number-density should come from the filament dynamics by comparing with the selection effects from the cylindrical filament and the density-redistribution introduced by RSD (see Appendix \ref{Appendix RSD Cylinder}). 

On the other hand, this infall velocity profile and the corresponding ``depletion radius'' in our cluster are very sensitive to the mass distribution in the system, including both the clusters and the filament. In observation, when increasing the mass at one end while reducing the mass at the other end, the projected infall flow at the more massive end will significantly increase (from $\sim20$ to $\sim60$ km/s), together with an increase in the depletion radius as this cluster becomes more dominant in the system (Fig.\,\ref{Fig unbalanced mass}). \textcolor{black}{Nonetheless, using our emulator constructed from N-body simulations (see Appendix \ref{Appendix Emulator} together with Fig.\,\ref{Fig filament model}, \ref{Fig model add RSD} \& \ref{Fig flow data sim}), we show step-by-step how the infall flow changes from a single halo case to a halo-halo model without a connecting filament, and then to halo-filament-halo scenarios, and finally add the RSD effect. We show the observed velocity profile generally agrees with our model, and is smaller than the halo-halo without filament case, being an order of magnitude smaller than the single halo case.}  This difference is attributed to the balance from the filament and the cluster at the other end, which significantly weakens the tidal gravitational field. The fact that a more massive environment can make the depletion radius shrink is also seen in N-body simulation \cite{Fong2021}. We additionally note that the balanced gravity from the environment is not the only reason the amplitude of our observed flow is smaller than the single halo case, as the excess redshift $\Delta z$ also removes the pairwise velocity of the two clusters.

\textcolor{black}{Although we validated that the observed excess redshift $\Delta z$ is velocity-originated, and the measured flow agrees with our model in terms of overall shape, we want to point out some potential imperfections in the current analysis that could lead to the difference in the amplitude of the flow. On the cosmology side, the fiducial cosmology of the group finder in observation (Planck2015) \cite{Tempel2017}, the simulation for the model (WMAP) \cite{Jing2019} and the mock catalog (Planck2015) \cite{Xu2024} in Appendix \ref{Appendix mock validation} we use could differ from the Universe. On the halo definition side, the group finder uses (friends-of-friends) FoF plus a group-galaxy distance-based selection comparing with $r_{200m}$ to account for merging halos \cite{Tempel2017}, while the simulation of our model uses FoF plus a halo-halo distance-based selection comparing with the depletion radius \cite{Han2018,Zhou2025}, and the mock catalog uses ROCKSTAR \cite{Behroozi2013}. The different halo definitions could not only lead to different halo masses, but also different density profiles and velocity profiles, including whether a galaxy is assigned to the cluster or the filament. Therefore, in order to directly derive filament mass from our measured velocity flow, one needs to consider all these effects in the simulation or mock catalog, which we leave for future explorations.}

\textcolor{black}{The observed excess redshift $\Delta z$ (Fig.\,\ref{Fig flow}) can be interpreted not only as velocity flow, but also as redshift-space distortion (RSD) \cite{Hamilton1998,Percival2009}, which suggests that the measured flow in Fig.\,\ref{Fig flow data sim} corresponds to the distorted density in Fig.\,\ref{Fig model add RSD}. The three stacked Kaiser pancakes in Fig.\,\ref{Fig model add RSD} further validate that the excess redshift $\Delta z$ is velocity-induced. Nonetheless, the wide errorbars in Fig.\,\ref{Fig flow} come from the fact that $\Delta z$ in Fig.\,\ref{Fig linear_fit} is widely distributed along the z-direction around the linear fit, which contains not only redshift measurement error, but also the Finger-of-God (FoG) effect and cosmic variance (including different filament mass distributions and velocity bulk flows).}

There are two extra phenomena we expect but do not observe due to insufficient S/N (signal-to-noise ratio) with a limited number of spectroscopic galaxies. One is the structure break at the center of the filament when the infall flow has existed for a sufficient duration: in Fig.\,\ref{Fig linear_fit} we see that the galaxy-line-density of the filament deviates from the single-peak distribution, however, the significance is low. But we do expect this phenomenon as the velocity in Fig.\,\ref{Fig flow data sim} and \ref{Fig filament model} suggests the outgoing flow overcomes the incoming flow of the filament right outside the filament center. The other is the possible tidal disruption of the less massive cluster if we keep increasing the mass-ratio in Fig.\,\ref{Fig unbalanced mass}, which could provide deeper insight into halo/cluster merger dynamics and offer valuable information on observational merger histories.

Previous studies have looked into cluster growth, cluster spins, filament spins independently \cite{Zhou2025,Goldstein2025,Gabriel-Silva2025,Wang2021}, while correlations between cluster spin and filament-cluster alignment have also been found \cite{Tang2025}. Our finding of how a filament falls into a cluster therefore serves as an important bridge connecting these previously disparate results. It also provides an independent view of how matter and angular momentum are transferred in such a complex non-linear environment.

In the near future, when more powerful spectroscopic data (DESI \cite{DESI2025DR2BAO}, PFS \cite{PFS2014}, MUST \cite{Zhao2024}, JUST \cite{JUST2024}, EAST \footnote{\href{https://astro.pku.edu.cn/research/PKU_EAST_ENG.htm}{PKU EAST Project Page}}, etc.) become available, one can further observe this filament flow and investigate its dependency on redshift, cluster mass and filament mass as a complete setup, which should provides strong S/N and opens a new window for examining gravity and constraining dark matter particle mass, as this flow is clearly sensitive to mass in the non-linear regime (Fig.\,\ref{Fig flow data sim}). \textcolor{black}{Nonetheless, the comparison with velocity reconstructions \cite{Wang2012,Yu2017,Yu2019,Xiao2025} could validate their accuracy on such non-linear scales, while the combination of our method and velocity reconstructions can help us remove the suppression from RSD and further improve the S/N.} Meanwhile, this flow can help in segmenting the observational data to trace the physical process of galaxies traveling through the filament-boundary-cluster timeline, and investigate properties like galaxy shape alignments and galaxy colors \cite{Yao2023,Yao2020}. By combining with SZ or X-ray observations, one can further understand the dynamical co-evolution of gas and structure formation.

\section*{Acknowledgements}

This work is supported by National Key R\&D Program of China No. 2022YFF0503403. 
JY acknowledges the support from NSFC Grant No.12203084 and 12573006. 
HYS acknowledges the support from NSFC of China under grant 11973070, the Shanghai Committee of Science and Technology grant No.19ZR1466600 and Key Research Program of Frontier Sciences, CAS, Grant No. ZDBS-LY-7013. 
PZ acknowledges the support of NSFC No. 11621303, the National Key R\&D Program of China 2023YFA1607800 and 2023YFA1607801.
P.W. is sponsored by Shanghai Rising-Star Program (No. 24QA2711100). 
This work is supported by the China Manned Space Program with grant no. CMS-CSST-2025-A03. 

JY and HX thank the ``Tree New Bee'' club for supporting friendly discussion environment. JY thanks DeepSeek \footnote{\url{https://chat.deepseek.com/}} for debugging, code parallelization, and language polishing.



\InterestConflict{The authors declare that they have no conflict of interest.}


\bibliographystyle{elsarticle-num} 
\bibliography{bibliography}

\begin{appendix}




\renewcommand{\thesection}{Appendix}

\section{Supplementary Materials}
\label{app1}

This appendix includes details for: \\
1. filament properties,\\
2. selection functions,\\
3. impact from nonlinear redshift-distance relation,\\
4. halo velocity/density profile emulators built from N-body simulation,\\
5. variations from fiducial filament model,\\
6. impact from redshift-space distortion (RSD) effect,\\
7. validation with mock catalog,\\
and 8. an illustration figure.

\subsection{Filament Properties} \label{Appendix Filament Properties}

\begin{figure}[H]
	\centering
	\includegraphics[width=0.5\textwidth]{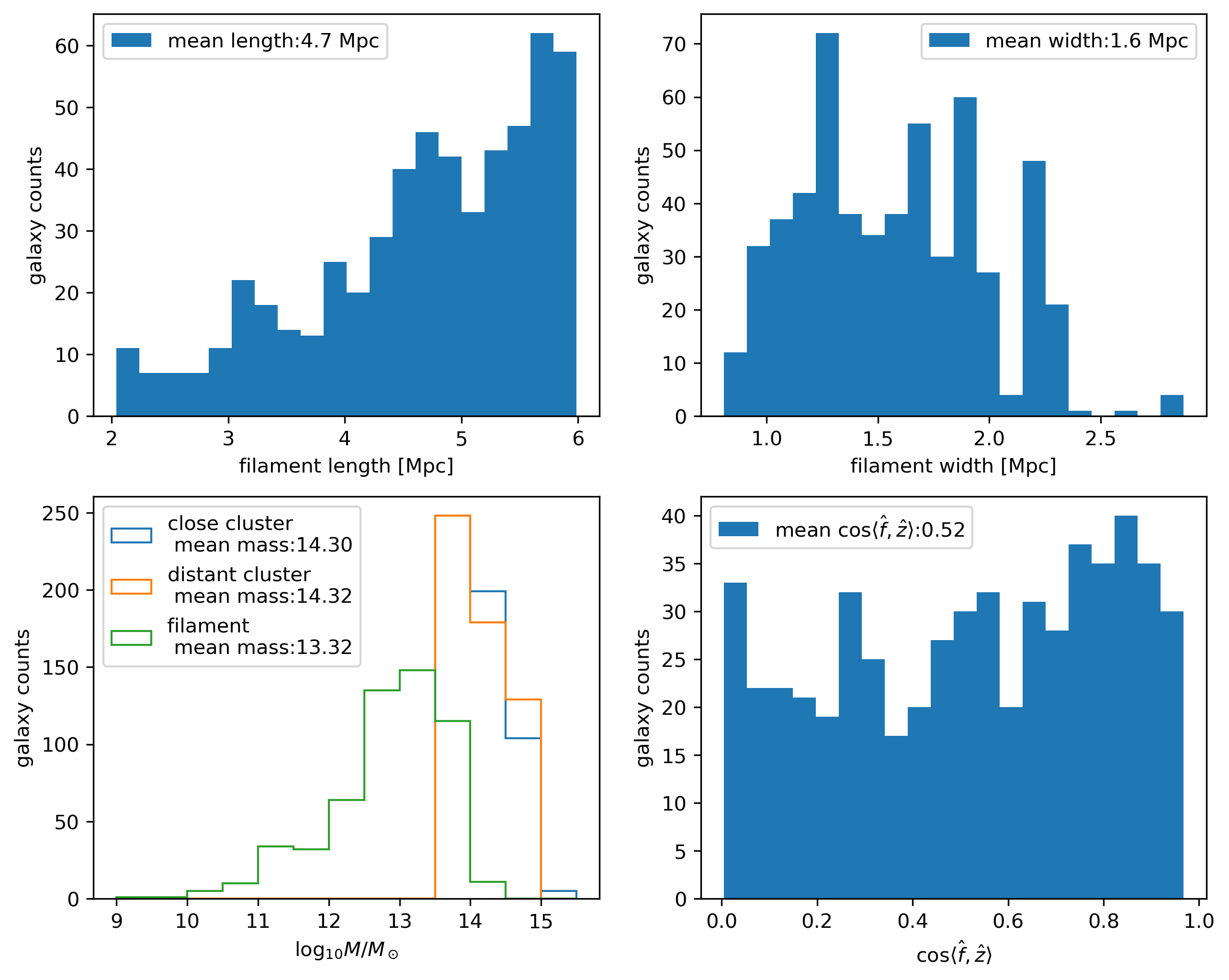}
	\caption{Filament properties (length, width, mass, alignment) of the selected cluster-filament-cluster system, with the fiducial selections described in the main text. By stacking the galaxies in each cluster-filament-cluster system, we find the mean length (cluster-cluster distance, top-left panel) is 4.7 Mpc, mean width (galaxy-filament axis distance, top-right panel) is 1.6 Mpc, mean filament mass is an order of magnitude smaller than the cluster mass (bottom-left panel), and mean alignment angle between the filament ($\hat{f}$, pointing from the low-z cluster to the high-z cluster) and the line-of-sight direction ($\hat{z}$) is $\sim60^\circ$ (bottom-right panel).}\label{Fig fiducial filament properties}
\end{figure}

Based on the fiducial selections of the cluster pairs and filaments described in the main text, we present the mean filament properties in this section. As the filament flow is actually described by each $\Delta z$ v.s. $x$ pair, we treat each galaxy as equally weighted in every cluster-galaxy-cluster system, and the corresponding distributions in terms of filament length, width, cluster mass, filament mass, and the filament alignment angle with respect to line-of-sight are presented in Fig.\,\ref{Fig fiducial filament properties}. Overall, we find the mean filament length is $L=4.7$ Mpc, width $W=1.6$ Mpc, cluster mass ${\rm log}_{10}M_c/M_\odot\sim14.3$, filament mass ${\rm log}_{10}M_f/M_\odot\sim13.32$, the filament alignment angle is ${\rm arccos}(0.52)\sim60^\circ$ with respect to the line-of-sight direction. We note the cluster mass and filament mass are calculated using the group mass of the designated galaxy. Since we require $\lambda_c>5$ for the cluster, its mass is relatively more reliable, while for the filament the selection of $\lambda_f<5$ could lead to extra bias\cite{Sun2022}. In the future, mass reconstructions with weak lensing for those systems could be very helpful.

\subsection{Selection Functions} \label{Appendix Functions}

\begin{figure}[H]
	\centering
	\includegraphics[width=0.5\textwidth]{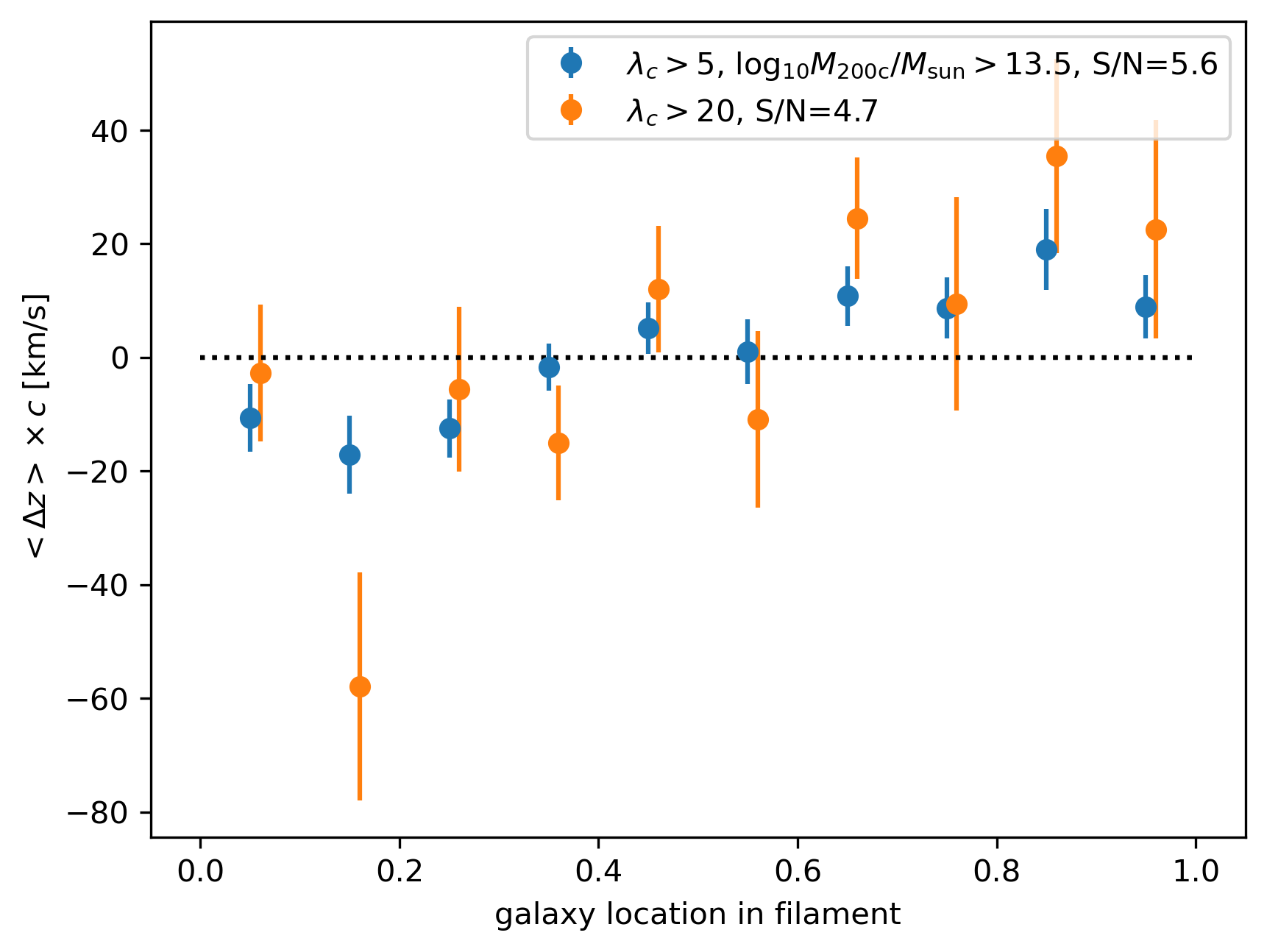}
	\caption{Comparison of different selections for the clusters. We see that pure richness selection (orange) can still yield a relatively strong S/N; however, its fitted slope has a significance of only $2\sigma$, due to the limited number of galaxies. Thus it is not an ideal selection for understanding detailed flow profile behavior.}\label{Fig cluster selection}
\end{figure}

A conventional choice for a cluster is richness $\lambda_c>20$; however, in SDSS galaxy group data only a small fraction of groups can satisfy this cut. Additionally, richness is not strongly correlated with mass, owing to the substantial scatter \cite{Murata2018}. We choose an alternative selection of $\lambda_c>5$ and $m>10^{13.5}M_\odot$, where $\lambda_c$ serves primarily to screen for reliable clusters \cite{Sun2022}, and use the optically obtained mass as the key selection. The comparison is shown in Fig.\,\ref{Fig cluster selection}.

\begin{figure}[H]
	\centering
	\includegraphics[width=0.5\textwidth]{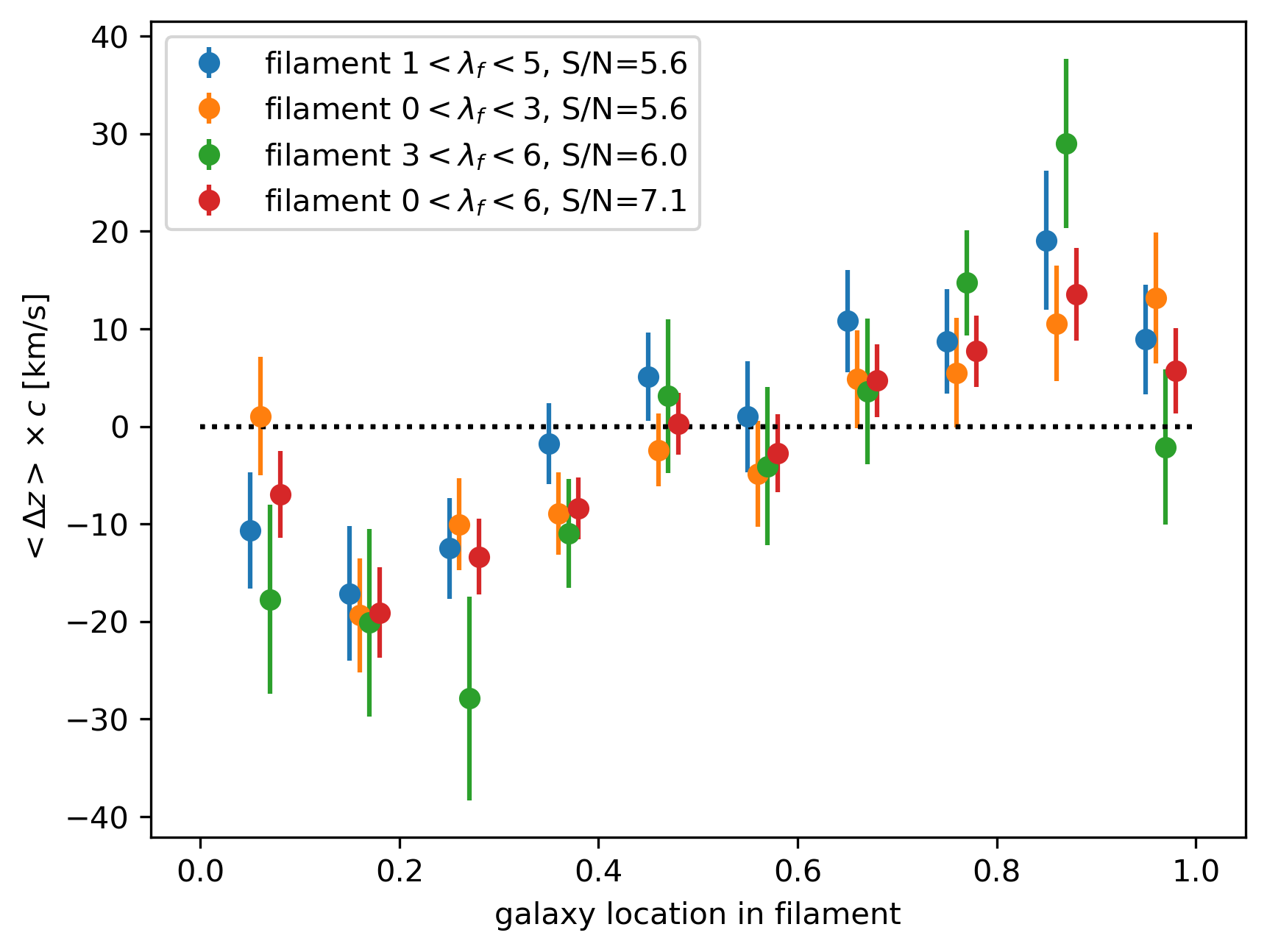}
	\caption{Comparisons of different filament richness cuts. The less obvious filaments ($0<\lambda_f<3$, orange) and more obvious filaments ($3<\lambda_f<6$, green) show significances similar to the default cut (blue), demonstrating that the filament flow widely exists. The aggressive cut ($0<\lambda_f<6$, red) can offer a more significant signal; however, we choose not to use it to ensure robustness.}\label{Fig filament richness}
\end{figure}

We check for the filament richness cut, in which a galaxy group not only must be located within a cylindrical space defined by the two clusters, but its richness must also satisfy $1<\lambda_f<5$. The choice of $\lambda_f<5$ is to distinguish filaments from the clusters selection $\lambda_c>5$, while the lower limit $\lambda_f>1$ is to minimize contamination from void regions where no filament exists between two clusters. Other choices of filament richness are shown in Fig.\,\ref{Fig filament richness}. We note that although the optimal case in terms of S/N ($\lambda_f<6$, S/N$\sim7$) uses a larger filament sample, we adopt the default cut, to maintain consistency with our cluster definition. 

\begin{figure}[H]
	\centering
	\includegraphics[width=0.5\textwidth]{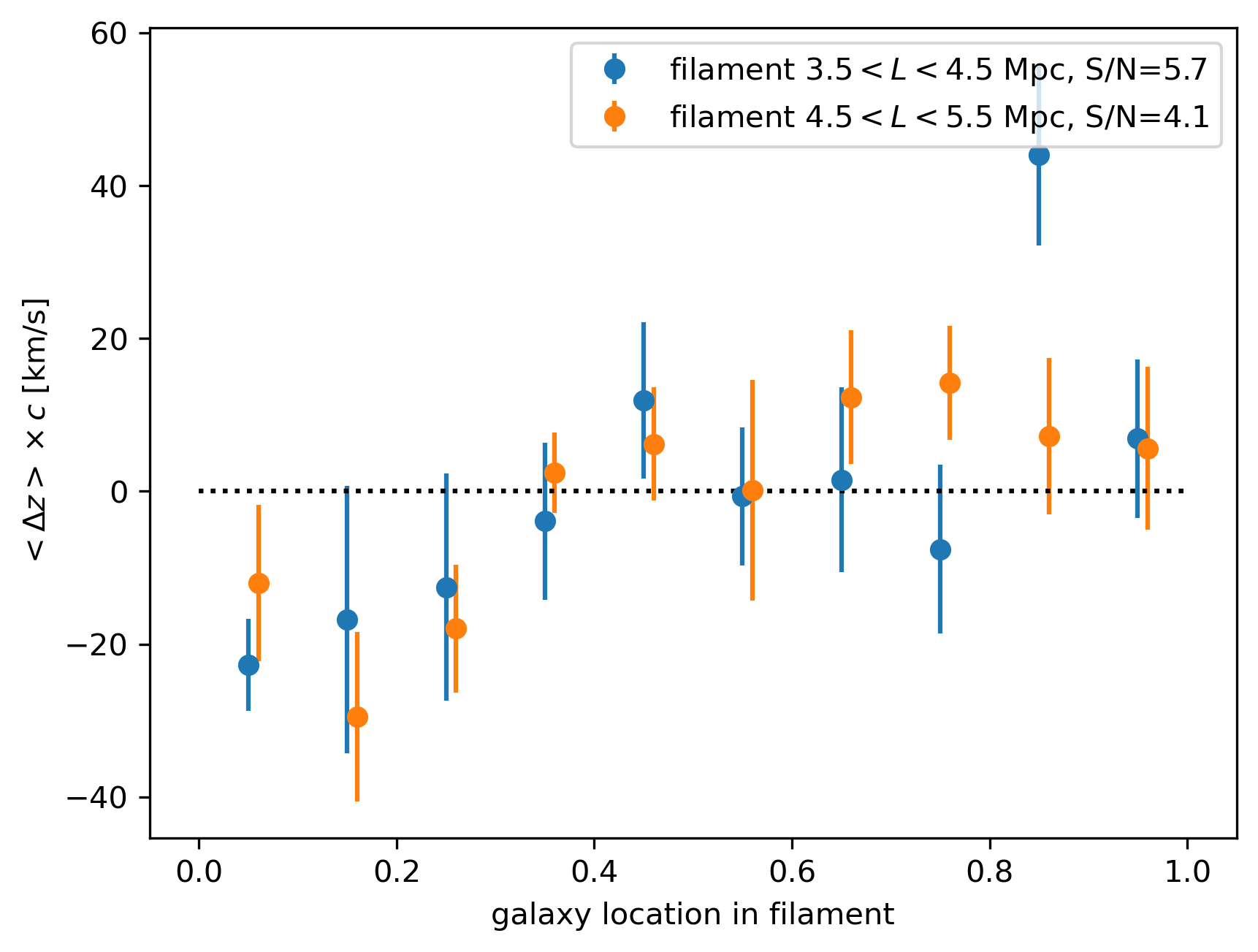}
	\caption{Comparisons of different filament length cuts. We find the majority of the filament flow signal comes from filament length $L\sim4$ and $\sim5$ Mpc. This is consistent with the general expectation for filament length. The longer or shorter filaments have a significance of $\le 3\sigma$, and we exclude them for clarity. The filament flow in this figure is not symmetric due to the limited number of galaxies.}\label{Fig filament length}
\end{figure}

Filament length is also an important property, as the flow velocity on a longer filament is lower due to weaker gravitational influence from the clusters, but a longer filament also means more member galaxies, which can reduce the noise. In Fig.\,\ref{Fig filament length}, the highest S/N is found for filament length $L\sim4$ and $\sim5$ Mpc, which falls within a reasonable range, as a quick calculation of the halo-matter correlation function\footnote{\url{https://github.com/czymh/csstemu} \cite{Chen2025}, more specifically \url{https://github.com/czymh/csstemu/blob/master/test/test-xihm.ipynb} } shows the 1-halo term is at $<1$ Mpc and the 2-halo term dominates around $\sim10$ Mpc, and we expect filaments to reside in between. 

\begin{figure}[H]
	\centering
	\includegraphics[width=0.5\textwidth]{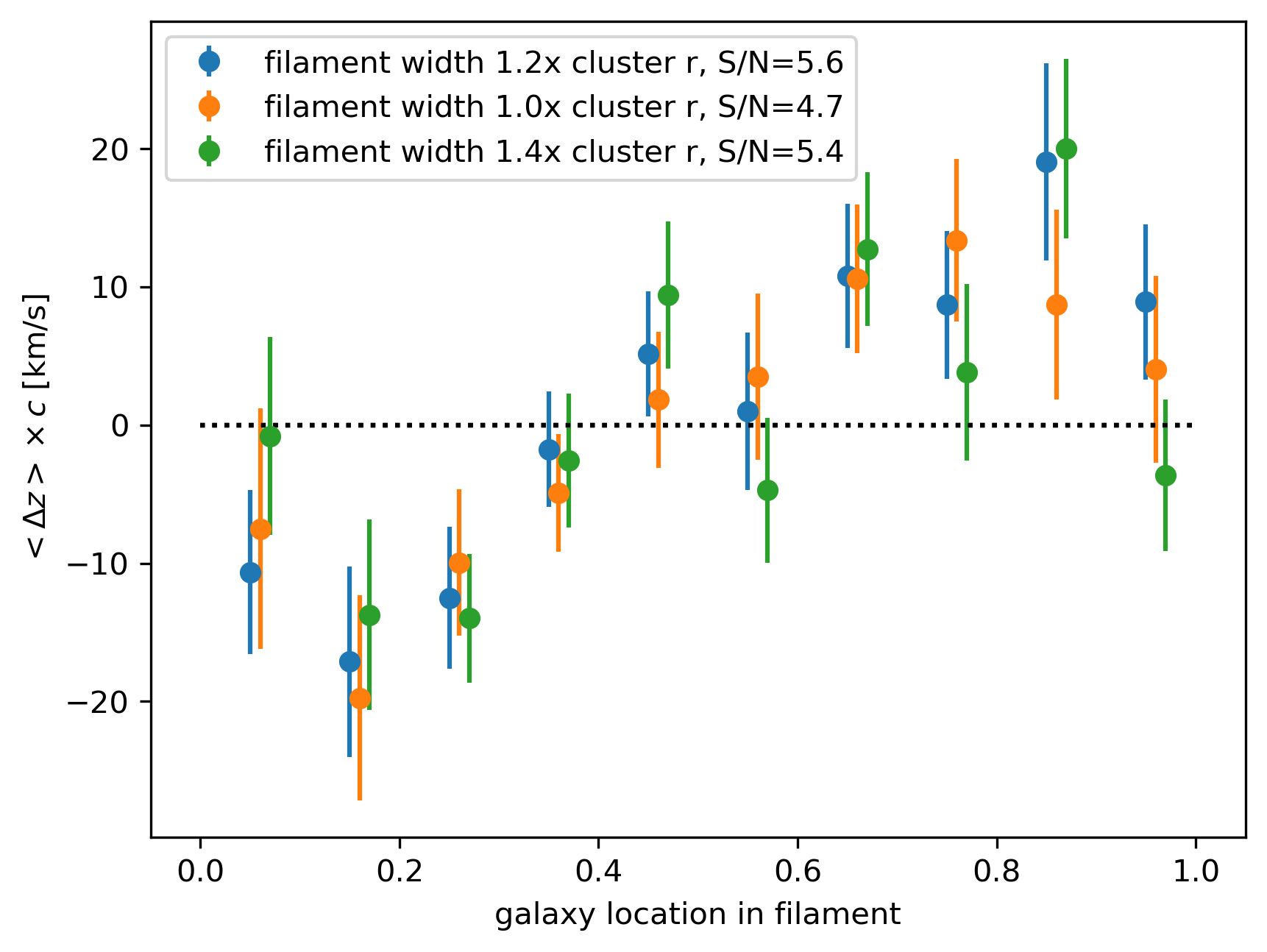}
	\caption{Comparisons of different choices for filament width. We see thinner filaments ($1\times r_{\rm 200c}$, orange) or fatter filaments ($1.4\times r_{\rm 200c}$, green) retain the main signal, but the fiducial choice ($1.2\times r_{\rm 200c}$, blue) optimizes the significance, likely because it accommodates the possible meandering paths or nonlinear geometry of filaments.}\label{Fig filament width}
\end{figure}

The last property we examine is the width of the filaments. We use $1.2\times r_{\rm 200c}$ as a fiducial choice, where $r_{\rm 200c}$ is the radius of the cluster where the mean density is 200 times higher than the critical density of the Universe. Considering that filaments are not perfectly straight lines between clusters and our cylindrical selection is only a first-order approximation, we use $1.2\times r_{\rm 200c}$ to include possible zigzag shapes of filaments. Comparisons with alternative widths can be found in Fig.\,\ref{Fig filament width}.

\subsection{\textcolor{black}{Impact from Nonlinear Redshift-Distance Relation}} \label{Appendix nonlinear d(z)}

\begin{figure}[H]
	\centering
	\includegraphics[width=0.5\textwidth]{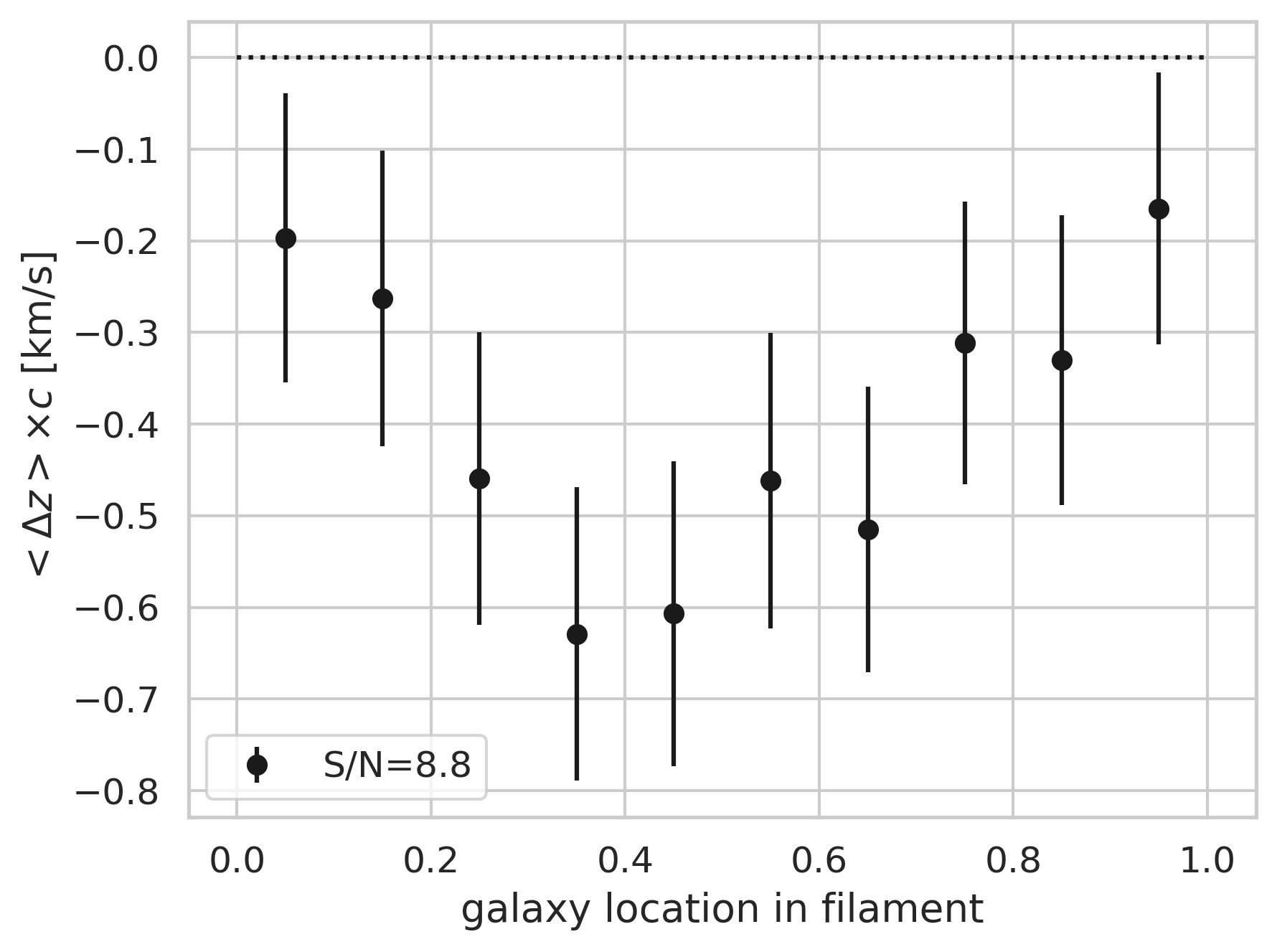}
	\caption{Excess redshift measurement with 1M randomly assigned dots in a cylindrical space that mimics the average filament properties. This proves our algorithm of linearly moving the rigid-body background redshift will only introduce a bias of $c\Delta z\le0.6$ km/s level, which is negligible compared with the measured signal.} \label{Fig nonlinear d(z)}
\end{figure}

To further validate the accuracy of linear interpolation and removal of the background redshift of our method, we randomly assign 1,000,000 dots in a cylindrical space that mimics our average porperties (i.e., z=0.06, L=4.7 Mpc, W=1.6 Mpc, aligned $60^\circ$ to the line-of-sight). The results are shown in Fig.\,\ref{Fig nonlinear d(z)}. Compared with the observation that peaks at $\sim20$ km/s, this $\le0.6$ km/s bias is negligible. For filaments with longer length, this bias could increase by a factor of 2 to 3; however, the flow signal could also increase significantly because the balance between the gravitational tidal fields from the clusters and the filament is changed, which allows for a faster infall flow. We leave this for future explorations with more specialized filament finders for long filaments.

\subsection{Emulator Build from N-body Simulation} \label{Appendix Emulator}

We construct a velocity emulator from N-body simulations \cite{Fong2021,Zhou2023,Zhou2025} to describe the general infall flow around clusters/halos, and use the emulator to construct a halo-halo-halo model to mimic the cluster-filament-cluster dynamics and compare it with the filament flow in this work. 

The simulation we use is one set from the CosmicGrowth Simulations \cite{Jing2019}, with a P$^3$M code \cite{Jing2002} run under WMAP $\Lambda$CDM cosmology ($\Omega_b=0.0445$, $\Omega_c=0.2235$, $\Omega_\Lambda=0.732$, $h=0.71$, $n_s=0.968$, and $\sigma_8=0.83$). The simulation box size is 600 Mpc/h with $3072^3$ dark matter particles, and the softening length is 0.01 Mpc/h. The halos are processed with Friends-of-Friends (FoF) and then with HBT+ \cite{Han2012,Han2018} to get subhalos and their evolutionary histories. The halo catalog covers a virial mass range $10^{11.5}<M_{\rm vir}[M_\odot/h]<3\times10^{15}$, where the minimum mass corresponding to $\sim500$ dark matter particles.

\begin{figure}[H]
	\centering
	\includegraphics[width=0.5\textwidth]{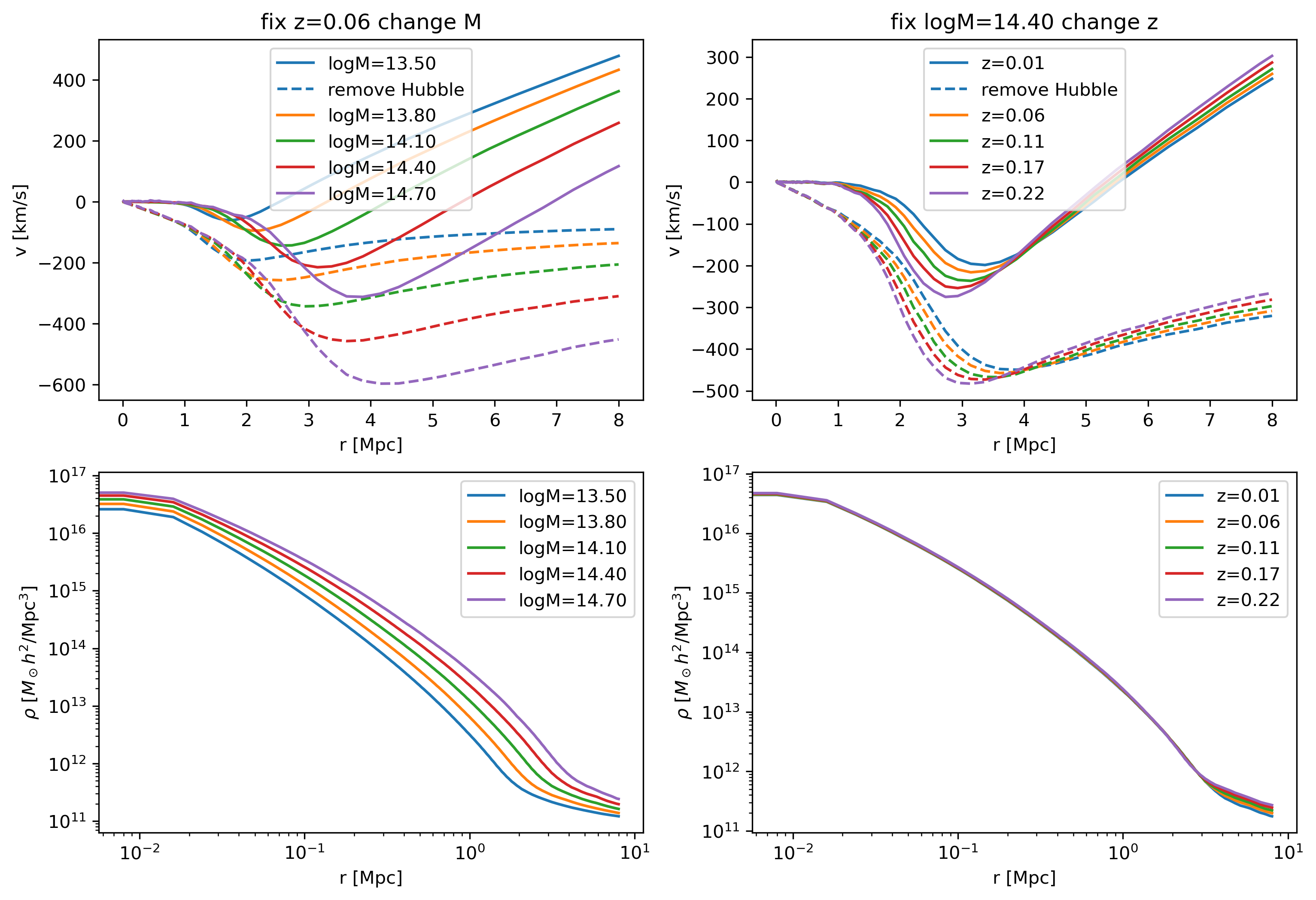}
	\caption{The emulator of the velocity profile and density profile of clusters/halos built from N-body simulations characterizing typical halo growth for the average cosmological environment. The top two panels show the velocity emulator $v(z,M,r)$, with the top-left panel fixing redshift while letting the mass change, and the top-right panel fixing the mass while letting the redshift change. Solid curves correspond to infall velocity plus Hubble flow, while dashed curves remove Hubble flow and leave only the infall velocity. The two bottom panels show the density emulator $\rho(z,M,r)$, with the bottom-left fixing $z$ and the bottom-right fixing $M$. We note the above emulators use virial mass $M_{\rm vir}$ instead of $M_{\rm 200c}$ in the data, and these mass definitions are linked by assuming a c-M relation \cite{Diemer2015}.}\label{Fig v emulator}
\end{figure}

To construct the emulator, we choose halos from 9 snapshots covering a redshift range $0<z<3$, and 9 mass bins covering $10^{12}<M_{\rm vir}[M_\odot/h]<10^{15}$. We linearly interpolate the velocity profile as a function of $z$, $\log_{10}(M_{\rm vir}/M_\odot)$, and comoving distance to the halo center $r$ [Mpc]. The resulting velocity emulator $v(z,M,r)$ is shown in the two top panels in Fig.\,\ref{Fig v emulator}. Similarly, the density emulator $\rho(z,M,r)$ is constructed and shown in the two bottom panels in Fig.\,\ref{Fig v emulator}. We additionally subtract the Hubble flow due to the expansion of the Universe, leaving the peculiar velocity profile of a given halo in a static frame.

\subsection{\textcolor{black}{Different Filament Models}} \label{Appendix Filament Model}

\begin{figure}[H]
	\centering
	\includegraphics[width=0.5\textwidth]{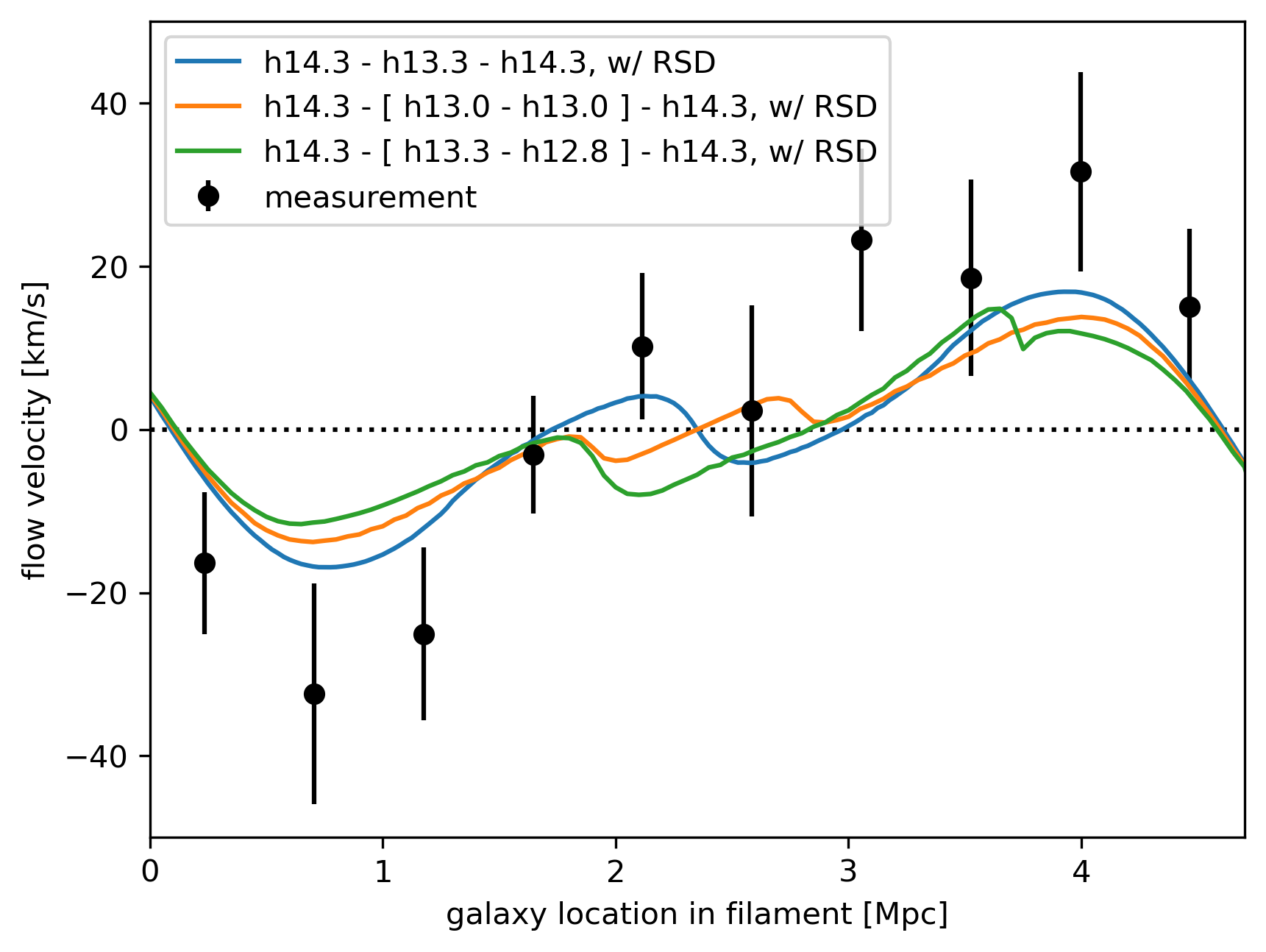}
	\caption{More complicated filament models. The black data dots are the measurements from observation, and the blue curve corresponds to our fiducial halo-halo-halo model, with a $M_{200c}=10^{13.3}M_\odot$ halo placed at the middle (loc=0.5) of the filament. We separated this halo into two smaller $10^{13.0}M_\odot$ halos and placed them symmetrically at loc=0.4 and loc=0.6, and the corresponding results are shown in orange. We also try an asymmetric model with a $10^{13.3}M_\odot$ halo at loc=0.4 and a $10^{12.8}M_\odot$ halo at loc=0.8, with the corresponding results shown in green.}\label{Fig different model}
\end{figure}

Aside from the demonstration in the main text that the flow is sensitive to the mass of clusters (Fig. 3) and the mass of the filament (Fig. 6), we further explore variations from the fiducial halo-halo-halo model to check the dependency on mass distribution along the filament. We explore two scenarios, both with two halos to represent the filament, while one model has symmetric mass distribution and the other is asymmetric. The results are shown in Fig.\,\ref{Fig different model}.

We see that each halo we put on the filament will cause a perturbation on the velocity of the flow at close range. Therefore, in the future when observational data increase significantly, the measurement could have a much better resolution to probe substructures along the filament.

\subsection{Details on the Redshift-Space Distortion (RSD) and Cylindrical Selection Function} \label{Appendix RSD Cylinder}

\begin{figure}[H]
	\centering
	\includegraphics[width=0.5\textwidth]{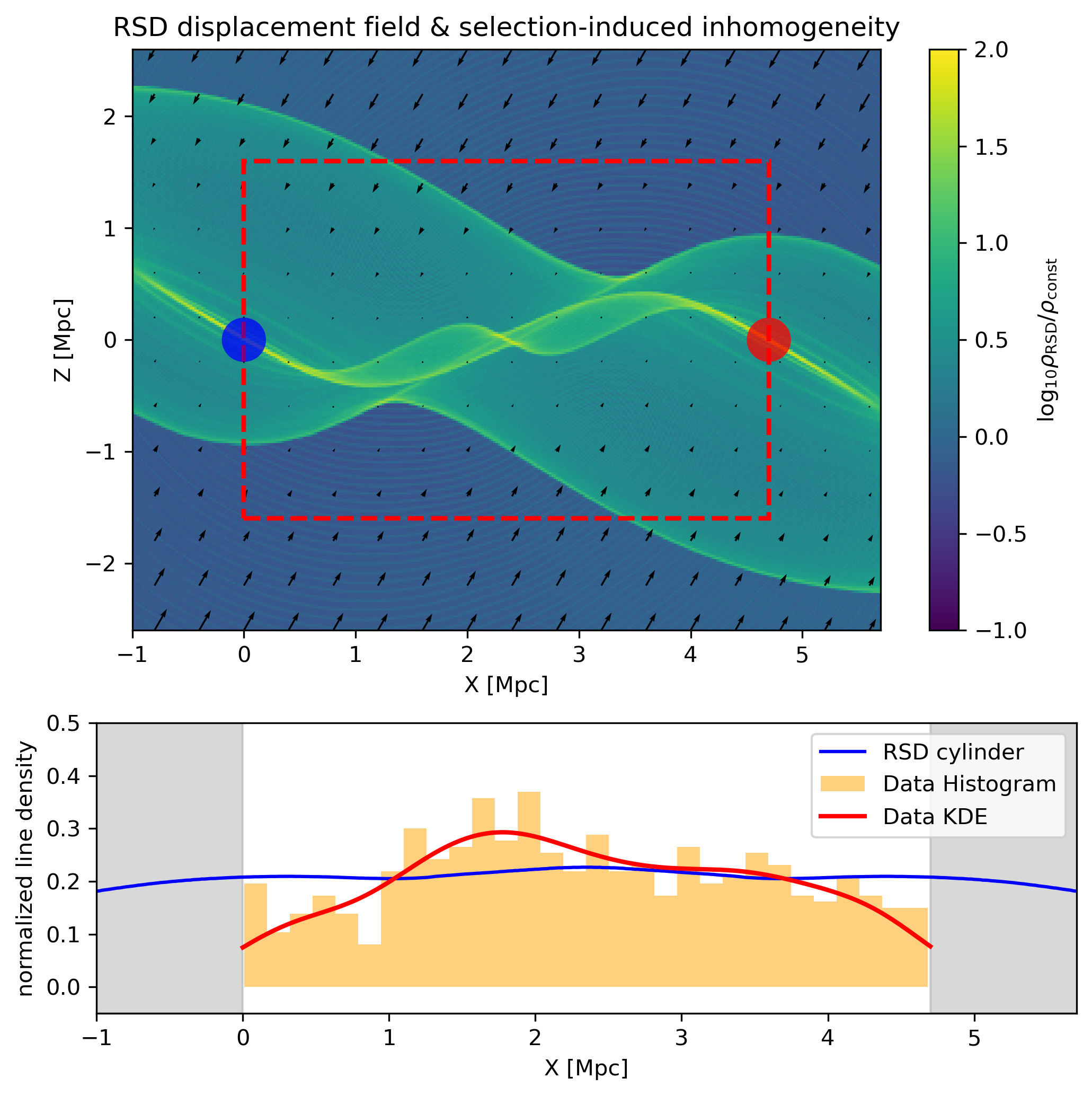}
	\caption{Impact from RSD and the cylindrical selection function. By constructing the cluster-filament-cluster model from simulation, we obtain the overall velocity field and project it onto the line-of-sight z-direction. This projected peculiar velocity of this special system induces a redshift change, and will form redshift-space distortion (RSD). In the top panel, we show the RSD displacement field in black arrows (with amplitude normalized to a visual-friendly amount). Assuming the real-space is filled with constant density $\rho_{\rm const}$, RSD will cause an inhomogeneous density distribution $\rho_{\rm RSD}$, which we also show in the top panel. We calculate the line-density along the X-axis within the cylindrical selection function (red dashed box) and show it as the blue curve in the bottom panel. \textcolor{black}{The galaxy distribution histogram in orange corresponds to Fig. 1 in the main text, and the red curve represents a smoothed distribution.} This demonstrates that the galaxy line-density in the main text is not due to the selection function, but due to the filament dynamics.}\label{Fig RSD cylinder}
\end{figure}

Nonetheless, we want to quantify the effect from RSD, given that the distance calculated in this work is in redshift space. The fact that the filament is aligned $\sim60^\circ$ to the z-direction makes the space distorted in a non-conventional way, unlike the typical Kaiser pancake or Finger-of-God shape. Moreover, the selection of cylindrical filament shape can affect the line-density. With the cluster-filament-cluster set-up in the main text, we can calculate the RSD displacement field and see how the real-space constant density distribution transfers to redshift-space. This effect is visualized in Fig.\,\ref{Fig RSD cylinder}, reflecting a twisted density distribution which is asymmetric around the X- or Z-axis. When the cylindrical filament is selected, its projected line-density on the X-axis is not a constant, shown in the bottom panel of Fig.\,\ref{Fig RSD cylinder}. We see that the galaxy line-density deviates from constant density in real-space, confirming that its dip-peak-dip feature is due to the filament dynamics but not RSD. More importantly, the RSD induced non-uniform density corresponds to the flow in the main text, but one visualized as RSD distance, while the other is visualized as velocity. This further confirms that the observed velocity flow is gravity-originated.

\subsection{\textcolor{black}{Validation with Mock Catalog}} \label{Appendix mock validation}

We apply our algorithm to the CCMD mock catalog \cite{Xu2024} to further validate our method. We note that the CCMD mock has different fiducial cosmology and halo definition from those in our observation and model; therefore, we only use it as a proof of principle, but not for direct comparison.

The CCMD mock catalog uses the MultiDark MDPL2 simulation, with a flat $\Lambda$CDM Planck2015 cosmology ($\Omega_m=0.307$, $\Omega_b=0.048$, $h=0.678$, $n_s=0.96$, $\sigma_8=0.823$). The simulation has a (1 Gpc/h)$^3$ box and 3840$^3$ particles, with mass resolution $1.51\times10^9~h^{-1}M_\odot$. The mock catalog is constructed based on a $z=0$ snapshot, and the halos are obtained by the phase-space halo finder ROCKSTAR \cite{Behroozi2013}.

\begin{figure}[H]
	\centering
	\includegraphics[width=0.5\textwidth]{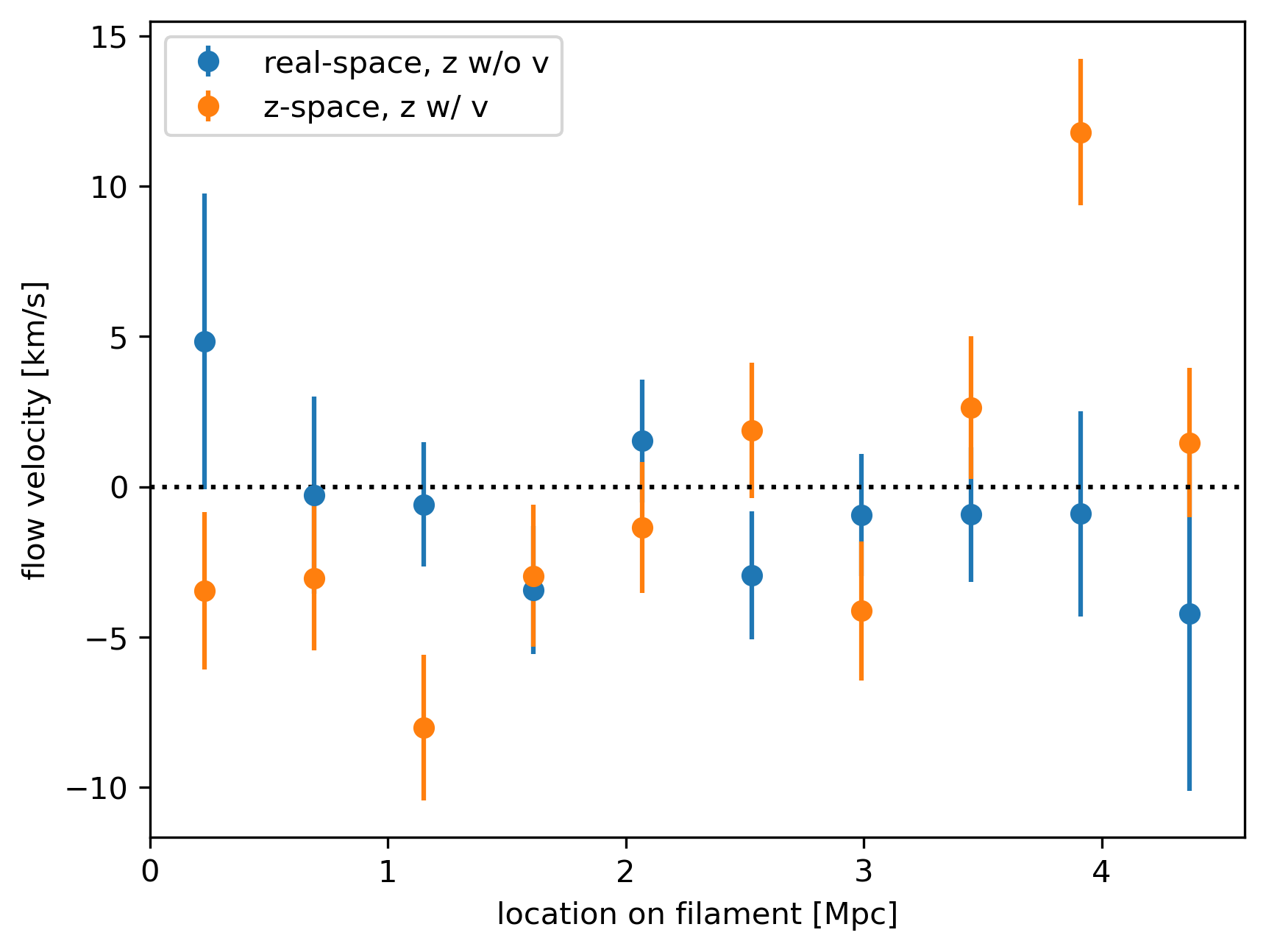}
	\caption{Validating the velocity flow measurement with CCMD mock catalog. When redshift is constructed only from the cosmological redshift but without contribution from peculiar velocity, the filament can be represented by a cylinder identified in real-space, and the corresponding results are in blue. When redshift is constructed from both the cosmological redshift and Doppler redshift, the filament is represented by a cylinder searched in redshift-space, and the corresponding results are in orange. We see the blue is consistent with 0, while the orange has a shape that is similar to our measurement and model in the main text. This further validates that our algorithm measures the velocity field. In addition, the number of galaxies in redshift-space is reduced by $\sim50\%$ compared to that in the real-space, demonstrating RSD also affects the assignment of galaxies into filament space.}\label{Fig CCMDbox mock}
\end{figure}

In Fig.\,\ref{Fig CCMDbox mock}, we see when the redshift is constructed only from the cosmological redshift, which means the catalog is treated as a static state, the output of our algorithm is consistent with 0. After adding the velocity contribution into redshift, the number of galaxies is reduced by $\sim50\%$. This suggests RSD has a strong impact on this algorithm, as demonstrated in Fig. 6 in the main text. Only when velocity's contribution is added in redshift as Doppler redshift, the output reveals an inflow shape, which is very similar to our results in the main text. This validates our approach successfully subtracts the velocity field information and breaks its degeneracy with cosmological redshift.

\subsection{\textcolor{black}{Illustration}} \label{Appendix illustration}

\begin{figure}[H]
	\centering
	\includegraphics[width=0.5\textwidth]{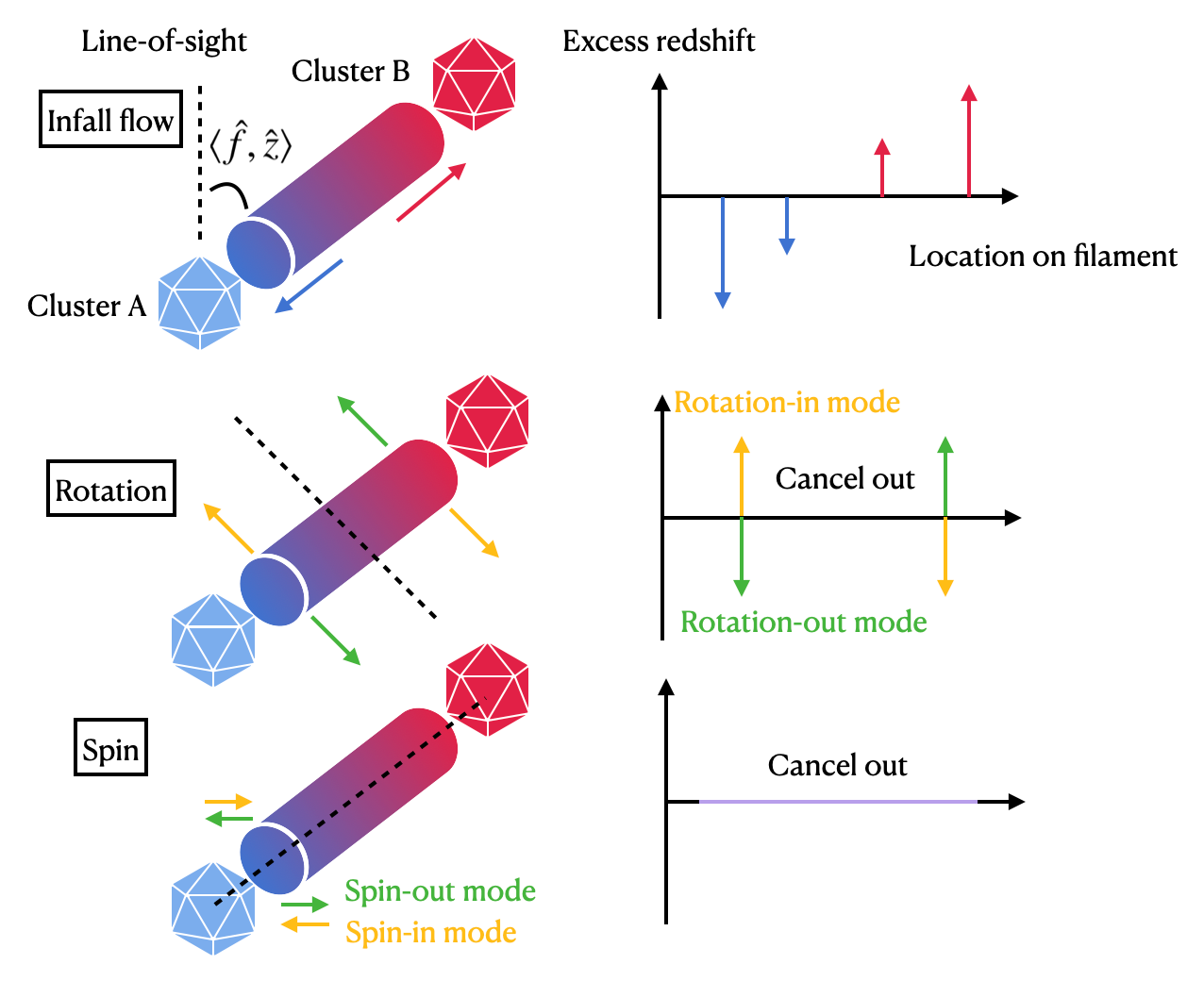}
	\caption{An Illustration of the cluster-filament-cluster set-up, and the comparison between different modes of infall velocity, rotation, and spin, with the induced excess redshift pattern.}\label{Fig illustration}
\end{figure}

In Fig.\,\ref{Fig illustration}, we illustrate the set-up of the cluster-filament-cluster system, and the difference between the infall flow mode and rotation and spin modes. Due to the cosmological principle, when stacking a large number of filaments, there should be an equal number of rotation-in and rotation-out modes, leading to an overall cancellation. Similarly, spin-in and spin-out modes should cancel out. Therefore, by stacking, the infall flow overwhelms the other modes.

\end{appendix}

\end{multicols}
\end{document}